\documentclass[journal,twocolumn]{IEEEtran}

\usepackage[colorlinks,urlcolor=blue,linkcolor=blue,citecolor=blue]{hyperref}
\usepackage{soul}
\usepackage{subcaption}
\usepackage{graphicx}
\usepackage{hyperref}
\usepackage{amsmath, amssymb,amsthm,cite, pict2e}
\usepackage{amsmath, amssymb,amsthm,cite}
\usepackage[utf8x]{inputenc}
\usepackage{placeins} 
\usepackage{graphicx}
\usepackage{psfrag}
\usepackage{bbm}
\usepackage{amsmath,amsfonts,amsthm,bm}
\usepackage{lipsum}
\usepackage{dblfloatfix}
\usepackage{mathtools}
\usepackage{algpseudocode}
\usepackage{algorithm}
\usepackage{color}
\usepackage{tcolorbox}
\usepackage{etoolbox}
\usepackage{physics}
\usepackage{tabularx}
\usepackage{multirow}
\usepackage{mleftright}

\allowdisplaybreaks
\newcommand{\RNum}[1]{\uppercase\expandafter{\romannumeral #1\relax}}
\usepackage{hyperref}
\usepackage{authblk}
\usepackage[english]{babel}
\usepackage{graphicx}
\usepackage{academicons}
 \usepackage{setspace}
\definecolor{orcidlogocol}{HTML}{A6CE39}

\newcommand{\indep}{\rotatebox[origin=c]{90}{$\models$}}
\usepackage{gensymb}

\begin{document}

\title{Neural Network-Based DOA Estimation in the Presence of Non-Gaussian Interference}

\author{S. Feintuch\thanks{Stefan Feintuch, Joseph Tabrikian, Igal Bilik, and Haim H. Permuter are with the School of Electrical and Computer Engineering, Ben Gurion University of the Negev, Beer Sheva, Israel. (e-mails: stefanfe@post.bgu.ac.il,  joseph@bgu.ac.il, bilik@bgu.ac.il,  haimp@bgu.ac.il). This work was partially
supported by the Israel Science Foundation under Grants 2666/19 and
1895/21.},
J. Tabrikian, {\it Fellow, IEEE}, I. Bilik, {\it Senior Member, IEEE}, and H. Permuter, {\it Senior Member, IEEE} 
}

\maketitle

\begin{abstract} 
This work addresses the problem of direction-
of-arrival (DOA) estimation in the presence of non-Gaussian,
heavy-tailed, and spatially-colored interference. Conventionally, the interference is considered to be Gaussian-distributed and spatially white. However, in practice, this assumption is not guaranteed, which results in degraded DOA estimation performance.
Maximum likelihood DOA estimation in the presence of non-Gaussian and spatially colored interference is computationally complex and not practical.
Therefore, this work proposes a neural network (NN) based DOA estimation approach for spatial spectrum estimation in multi-source scenarios with {\it a-priori} unknown number of sources in the presence of non-Gaussian spatially-colored interference. 
The proposed approach utilizes a single NN instance for simultaneous source enumeration and DOA estimation.
It is shown via simulations that the proposed approach significantly outperforms conventional and NN-based approaches in terms of probability of resolution, estimation accuracy, and source enumeration accuracy in conditions of low SIR, small sample support, and when the angular separation between the source DOAs and the spatially-colored interference is small.


\end{abstract}

\section{Introduction}\label{sec:introduction}

Direction-of-arrival (DOA) estimation using a sensor array is required in multiple applications, such as radar, sonar, ultrasonic, wireless communications, and medical imaging~\cite{van2004optimum}. 
In real-world applications, the signal received at the sensor array is a superposition of signals from the sources of interest, interference, and receiver thermal noise. In radars, the received signal consists of a target echo, clutter, and thermal noise. 
In multiple scenarios, the radar clutter has a spatially-colored, heavy-tailed non-Gaussian distribution~\cite{ollila2012complex}, which can significantly degrade the performance of conventional estimators.

Minimum-variance-distortionless-response (MVDR)~\cite{capon1969high}, is a conventional adaptive beamforming approach for DOA estimation. 
MVDR estimates the spatial spectrum and obtains the source DOAs via a one-dimensional peak search on a predefined grid.
The estimation of signal parameters using rotational invariance techniques (ESPIRIT)~\cite{roy1989esprit}, multiple signal classification (MUSIC)~\cite{schmidt1986multiple}, and root-MUSIC (R-MUSIC)~\cite{barabell1983improving} are additional widely used DOA estimation approaches. 
These approaches involve received signal autocorrelation matrix processing, which conventionally is performed via the sample autocorrelation matrix estimation~\cite{capon1969high,schmidt1986multiple,roy1989esprit,barabell1983improving}. 
However, the performance of the sample autocorrelation matrix estimator degrades in small sample support or non-Gaussian scenarios. 
Furthermore, these methods use second-order statistics only and omit the higher-order statistics on non-Gaussian-distributed interference.
In addition, ESPRIT, MUSIC, and R-MUSIC approaches require {\it a-priori} knowledge of the number of sources (or targets), which limits their practical use.

The problem of DOA estimation in the presence of non-Gaussian interference is of great practical interest. 
The maximum likelihood estimator (MLE) for  DOA estimation in the presence of non-Gaussian interference does not have a closed-form analytical solution~\cite{besson2016direction,singh2021kernel}.
Multiple model-based DOA estimation approaches have been intensively studied in the literature~\cite{besson2016direction,singh2021kernel,ollila2008influence,fortunati2019semiparametric,luan2021generalized,todros2015robust,yazdi2020measure,zhang2016maximum,zhang2017mimo,meriaux2019iterative,trinh2021partially, dai2017sparse}.

Robust covariance matrix-based DOA estimation and source enumeration methods have been studied in the literature.
For complex elliptically symmetric (CES) distributed data, the authors in \cite{ollila2008influence} showed that a scatter matrix-based beamformer is consistent, and the semiparametric lower bound, and Slepian-Bangs formula for DOA estimation were derived in \cite{fortunati2019semiparametric}.
In~\cite{luan2021generalized}, a generalized covariance-based (GC) approach for the covariance matrix estimation in scenarios with impulsive alpha-stable noise was proposed for MUSIC DOA estimation. 
However, these methods consider a specific family of distributions, such as the CES or alpha-stable, and are, therefore, limited in the case of model mismatch.
In~\cite{todros2015robust}, a probability measure transform (MT) based covariance matrix estimator was proposed for MUSIC-based DOA estimation and minimum descriptive length (MDL) based source enumeration.
The MT-based covariance estimator was also adopted for robust MVDR beamformer~\cite{yazdi2020measure}.
These methods are usually based on setting a parameter that determines the tradeoff between the level of robustness and performance.

The problem of DOA estimation in the presence of a mixture of spatially-white K-distributed and Gaussian-distributed noise under a deterministic and unknown (conditional) source model was studied in~\cite{besson2016direction}.
An iterative MLE-based approach for the conditional and joint likelihood of interference distribution's parameters was derived in~\cite{zhang2016maximum,zhang2017mimo}.
This approach was further extended in~\cite{meriaux2019iterative} to  the marginal likelihood function.
However, this approach is computationally complex due to numerical integral evaluation that involves a $2M$ dimensional grid search for $M$ targets \cite{singh2021kernel}. Therefore,~\cite{singh2021kernel} proposed a kernel minimum error entropy-based adaptive estimator and a novel criterion to reduce the estimator's computational complexity. 
The expectation-maximization (EM) with a partial relaxation-based DOA estimation algorithm under the conditional model assumption was proposed in~\cite{trinh2021partially}. 
In \cite{dai2017sparse}, a sparse Bayesian learning (SBL) approach for outlier rejection of impulsive and spatially-white interference was proposed. This EM-based approach did not require {\it a-priori} knowledge of the number of sources and was shown to resolve highly-correlated and coherent sources.
However, none of these model-based DOA estimation approaches considered an \textit{a-priori} an unknown number of sources and spatially-colored interference and therefore are limited for real-world applications.
Although source enumeration methods, such as MDL and Akaike information criterion (AIC)~\cite{wax1985detection} can be used, they assume signal Gaussianity and can therefore be inaccurate in non-Gaussian scenarios.

Deep learning and machine learning approaches were recently adopted for radar signal processing.
Three types of NN-based DOA estimation approaches have been introduced in literature~\cite{fuchs2022machine}.
The first approach assumes \textit{a-priori} known number of sources and uses a NN, which is optimized to output a vector of the estimated DOAs~\cite{fuchs2019single,el1997performance,milovanovic2012application,ofek2011modular,barthelme2021machine,bialer2019performance,cong2020robust}.
The second approach does not assume \textit{a-priori} known number of sources and uses a NN for source enumeration~\cite{barthelme2021machine,bialer2019performance,cong2020robust,gardill2018multi,fuchs2019model,rogers2019estimating,rogers2021robust}.
The third approach uses a NN to estimate source presence probability at each DOA on a predefined angular grid and obtains the source DOAs via a peak search~\cite{liu2018direction,ozanich2019deep,gall2020spectrum,gall2020learning,ahmed2020deep,papageorgiou2020direction,papageorgiou2021deep,ozanich2020feedforward,yao2020crnn,barthelme2021doa}.
However, all these approaches have not addressed non-Gaussian and spatially-colored interference~\cite{fuchs2022machine,fuchs2019single,el1997performance,milovanovic2012application,ofek2011modular,barthelme2021machine,bialer2019performance,cong2020robust,gardill2018multi,fuchs2019model,rogers2019estimating,rogers2021robust,liu2018direction,ozanich2019deep,gall2020spectrum,gall2020learning,ahmed2020deep,papageorgiou2020direction,papageorgiou2021deep,ozanich2020feedforward,yao2020crnn,barthelme2021doa}.

The cases of non-Gaussian and/or spatially-colored interference have been addressed using machine learning-based approaches.
For massive MIMO cognitive radar, a reinforcement learning-based approach for multi-target detection under heavy-tailed spatially-colored interference was proposed in \cite{ahmed2021reinforcement}.
In~\cite{luo2022deep}, authors addressed the MIMO radar target detection under non-Gaussian spatially-colored interference by using a CNN architecture that is optimized according to a novel loss.
A radial-basis-function (RBF) NN~\cite{guo2008performance} and a convolutional neural network (CNN)~\cite{chen2020novel} architectures were proposed for DOA estimation in the presence of non-Gaussian impulsive noise. In~\cite{chen2022multisource}, a CNN-based architecture that includes denoising NN, source enumeration NN, and DOA estimation sub-NNs, was introduced.  
However, \cite{guo2008performance,chen2020novel,chen2022multisource} consider spatially-white noise and are suboptimal for scenarios with spatially-colored interference.

\textcolor{black}{In \cite{feintuch2022neural}, a novel NN-based approach is introduced for radar target detection under slow-time correlated and heavy-tailed distributed clutter in the range-Doppler domain without any spatial processing.
This work extends \cite{feintuch2022neural} to the field of array processing, by employing a similar NN-based approach under performance criteria for source DOA estimation and enumeration. 
More specifically, we address the array processing problem with \textit{a-priori} unknown number of sources in the presence of non-Gaussian, heavy-tailed, spatially-colored interference at a low signal-to-interference ratio (SIR) and small sample size.}
The contribution of this work includes:
\begin{enumerate}
    \item A novel NN-based processing mechanism is used for 
    array processing within non-Gaussian spatially-colored interference. The proposed NN architecture utilizes the structure of information within the set of received complex snapshots.
    \item The proposed NN is optimized to output an interference-mitigated spatial spectrum and is used for simultaneous source enumeration and DOA estimation of sources within non-Gaussian spatially-colored interference.
\end{enumerate}
The proposed approach outperforms conventional adaptive beamforming and competes for straightforward NN-based methods in terms of probability of resolution and estimation accuracy in scenarios with non-Gaussian spatially-colored interference.
In addition, the proposed approach outperforms conventional source enumeration techniques in scenarios characterized by non-Gaussian spatially-colored interference.

The following notations are used throughout the paper. Roman boldface lower-case and upper-case letters represent vectors and matrices, respectively, while Italic letters stand for scalars. 
$\mathbf{I}_N$ is the identity matrix of size $N \times N$ and   $\mathbf{1}_N$ is a column vector of length $N$ whose entries are equal to one.
$\mathbb{E}(\cdot)$, $(\cdot)^T$, and $(\cdot)^H$ are the expectation, transpose, and Hermitian transpose operators, respectively.
$\text{Vec}(\cdot)$, $\text{diag}(\cdot)$, and $|\cdot|$ stand for the vectorization, diagonalization, and absolute value operators, respectively.
$[\mathbf{a}]_{n}$ and $[\mathbf{A}]_{n,m}$ are the $n$-th and $n,m$-th elements of the vector $\mathbf{a}$ and the matrix $\mathbf{A}$, respectively.

The remainder of this paper is organized as follows. The addressed problem is stated in Section \ref{sec:problem_statement}. Section \ref{sec:proposed_method} introduces the proposed NN-based DOA estimation approach. 
The proposed approach is evaluated via simulations in Section \ref{sec:eval}. Our conclusions are summarized in Section \ref{sec:conclusion}.

\section{Problem Definition}\label{sec:problem_statement}
% The addressed problem is presented in this Section.
% The measurement model is described in Subsection~\ref{subsec:meas_model}, and the conventional adaptive beamforming method, is briefly presented in Subsection~\ref{subsec:mvdr}.

% \subsection{Measurements Model}\label{subsec:meas_model}
This work considers the problem of DOA estimation using an array  of $L$ receiving elements and $M$ distinct and unknown sources with DOAs, $\bm{\Theta}=\{\theta_1,\dots,\theta_M\}$.
The measurements contain $K$ spatial snapshots, $\{\mathbf{x}_k\}_{k=1}^K$:
% The spatial $k$-th snapshot, $\mathbf{x}_k$, is modeled as:
\begin{align}\label{eq:meas_model}
    \mathbf{x}_k &= \mathbf{A}\left(\bm{\Theta}\right)\mathbf{s}_k + \sigma_c\mathbf{c}_k + \mathbf{n}_k\;,\\ 
    &=\sum_{m=1}^{M}{\mathbf{a}\left(\theta_m\right)s_{k,m}} + \sigma_c\mathbf{c}_k + \mathbf{n}_k\;,\;k=1,\dots,K\;,\nonumber
\end{align}
where $\mathbf{A}\left(\bm{\Theta}\right)=\begin{bmatrix}\mathbf{a}\left(\theta_1\right)&\cdots&\mathbf{a}\left(\theta_M\right)\end{bmatrix}$, with $\mathbf{a}\left(\theta_m\right)\in\mathbb{C}^{L}$ denoting the steering vector for source at direction $\theta_m$, and $\mathbf{s}_k\triangleq\begin{bmatrix}s_{k,1}&\cdots&s_{k,M}\end{bmatrix}^T$ is the source signal vector.
We assume an unconditional model~\cite{stoica1990performance}, where  $\{\mathbf{s}_k\}\stackrel{i.i.d.}{\sim}\mathcal{CN}\left(\mathbf{0}_M,\text{diag}\left(\sigma_1^2,\dots,\sigma_M^2\right)\right)$, is temporally uncorrelated between pulses. The targets are assumed to be spatially distinct.
The receiver thermal noise, denoted by  $\mathbf{n}_k$, is considered to be complex Gaussian-distributed $\{\mathbf{n}_k\}\stackrel{i.i.d.}{\sim}\mathcal{CN}\left(\mathbf{0}_L,\sigma_n^2\mathbf{I}_L\right)$.
The heavy-tailed non-Gaussian and spatially-colored interference is modeled by the interference amplitude $\sigma_c$, and the interference component $\mathbf{c}_k\in\mathbb{C}^L$.
The considered compound-Gaussian distributed interference, $\{\mathbf{c}_k\}\stackrel{i.i.d.}{\sim}\mathcal{K}\left(\nu,\theta_c\right)$ represents a non-Gaussian interference with angular spread around an unknown direction $\theta_c$, such that $\mathbf{c}\sim\mathcal{K}\left(\nu,\theta_c\right)$ implies
    \begin{align}\label{eq:interf_model}
        &\mathbf{c}=\sqrt{\tau}\mathbf{z}\;,\\
        &\tau\indep\mathbf{z},\;\tau\sim\Gamma\left(\nu,\nu\right),\;\mathbf{z}\sim\mathcal{CN}\left(\mathbf{0}_L,\mathbf{M}_{\theta_c}\right)\;.\nonumber
    \end{align}
The compound-Gaussian statistical model is conventionally used in the literature to model heavy-tailed non-Gaussian interference~\cite{singh2021kernel,zhang2016maximum,meriaux2019iterative,besson2016direction,viberg1997maximum,luo2022deep}.
The texture component, $\tau\in\mathbb{R}_{+}$, determines the heavy-tailed behavior and is characterized by $\nu$.
The speckle component, $\mathbf{z}\in\mathbb{C}^L$, determines the spatial distribution of the interference and is characterized by the covariance matrix, $\mathbf{M}_{\theta_c}$.
The spatial covariance matrix of the interference upholds:
\begin{align}\label{eq:interf_cov_mat}
    \mathbb{E}\left[\sigma_c^2\mathbf{c}\mathbf{c}^H\right]=&\sigma_c^2\mathbb{E}\left[\tau\right]\mathbb{E}\left[\mathbf{z}\mathbf{z}^H\right]=\sigma_c^2\mathbf{M}_{\theta_c}\;,
\end{align}
where $\mathbf{M}_{\theta_c}$ can be modeled as~\cite{zhang2016maximum,zhang2017mimo,meriaux2019iterative,viberg1997maximum,luo2022deep}:
\begin{align}\label{eq:M_theta_c}
    \left[\mathbf{M}_{\theta_c}\right]_{m,l}&=\rho^{\left|m-l\right|}e^{j(m-l)\pi\sin\theta_c}\;.
\end{align}
The model in \eqref{eq:interf_cov_mat} and \eqref{eq:M_theta_c}, represents the spatial interference, characterized by $\rho$, with a spread around the interference DOA, $\theta_c$.

% The interference-to-noise ratio (INR) and signal-to-interference ratio of the $m$-th source, $\text{SIR}_m$, are defined as:
% \begin{align}
%     \text{INR}&=\frac{\mathbb{E}\left[\left\|\mathbf{c}\right\|^2\right]}{\mathbb{E}\left[\left\|\mathbf{n}\right\|^2\right]}=\frac{\sigma_c^2}{\sigma_n^2}\;,\\
%     \text{SNR}_m&=\frac{\mathbb{E}\left[\left\|\mathbf{a}\left(\theta_m\right)s_m\right\|^2\right]}{\mathbb{E}\left[\left\|\mathbf{n}\right\|^2\right]}=\frac{\sigma_m^2}{\sigma_n^2}\;,\nonumber\\
%     \text{SIR}_m&=\frac{\mathbb{E}\left[\left\|\mathbf{a}\left(\theta_m\right)s_m\right\|^2\right]}{\mathbb{E}\left[\left\|\mathbf{c}\right\|^2\right]}=\frac{\sigma_m^2}{\sigma_c^2}\;.\nonumber
% \end{align}

\section{The Proposed DAFC-Based Neural Network}\label{sec:proposed_method}
% The proposed NN-based approach for DOA estimation is detailed in this Section. 
% It is proposed to incorporate both the MOS and regression methods to generate the number of peaks in the DOA spectrum. 
% This approach jointly outputs the MOS for source enumeration and the vector of estimated DOAs. 

In the following, the data pre-processing and the proposed NN-based processing mechanism are introduced in Subsections~\ref{subsec:pre_processing} and \ref{subsec:dafc}. The proposed NN architecture and loss function are detailed in Subsections~\ref{subsec:nn_arch} and \ref{subsec:nn_training}, respectively.

The proposed approach is \textcolor{black}{ similar to} the NN architecture that was introduced for linear-frequency-modulated (LFM) radar target detection in the range-Doppler domain~\cite{feintuch2022neural}.
\textcolor{black}{The proposed architecture in this work utilizes the dimensional alternating fully-connected (DAFC) mechanism introduced in \cite{feintuch2022neural}.
However, since this work addresses a different problem, we have empirically found that the pre-processing, NN architecture, and loss function described in Subsections~\ref{subsec:pre_processing}, \ref{subsec:nn_arch}, and \ref{subsec:nn_training}, respectively, allow to attain better results.}

\subsection{Pre-Processing}\label{subsec:pre_processing}
The input matrix, $\mathbf{X}\in\mathbb{C}^{L\times K}$ is constructed from the set of $K$ snapshots in~\eqref{eq:meas_model}, $\{\mathbf{x}_k\}$:
\begin{align}\label{eq:input_matrix_X}
    &\mathbf{X} = \begin{bmatrix} \mathbf{x}_1 & \mathbf{x}_2 & \cdots & \mathbf{x}_K \end{bmatrix}\;,
\end{align}
where the $k$-th column of $\mathbf{X}$ contains the $k$-th snapshot.
The variation between the columns of $\mathbf{X}$ is induced by the statistical characteristics of the source signal $\mathbf{s}_k$, interference signal $\mathbf{c}_k$, and thermal noise $\mathbf{n}_k$.
Therefore, each column in $\mathbf{X}$ can be interpreted as a complex ``feature'' vector containing essential information for DOA estimation. The set of columns in $\mathbf{X}$ can be interpreted as ``realizations'' of that feature.

The complex-valued matrix, $\mathbf{X}$, is converted into a real-valued representation needed for the NN-based processing. 
To keep consistency with~\cite{feintuch2022neural}, we apply a transpose operator to the input matrix, such that the snapshots are stacked in rows.
The output of the pre-processing denoted by $\mathbf{Z}_0\in\mathbb{C}^{K\times 2L}$, is:
\begin{align}\label{eq:Z_0}
    \mathbf{Z}_0 = \left[\Re{\mathbf{X}^T},\;\Im{\mathbf{X}^T}\right]\;.
\end{align}
% Notice that the concept of ``features'' and ``realizations'' in $\mathbf{X}$ is valid in $\mathbf{Z}_0$ since the spectral content of the real and imaginary parts of each row in $\mathbf{Z}_0$ is similar to that of each column in $\mathbf{X}$.

\subsection{Dimensional Alternating Fully-Connected}\label{subsec:dafc}

\begin{figure}[t]
\centering
    \includegraphics[width=0.45\textwidth]{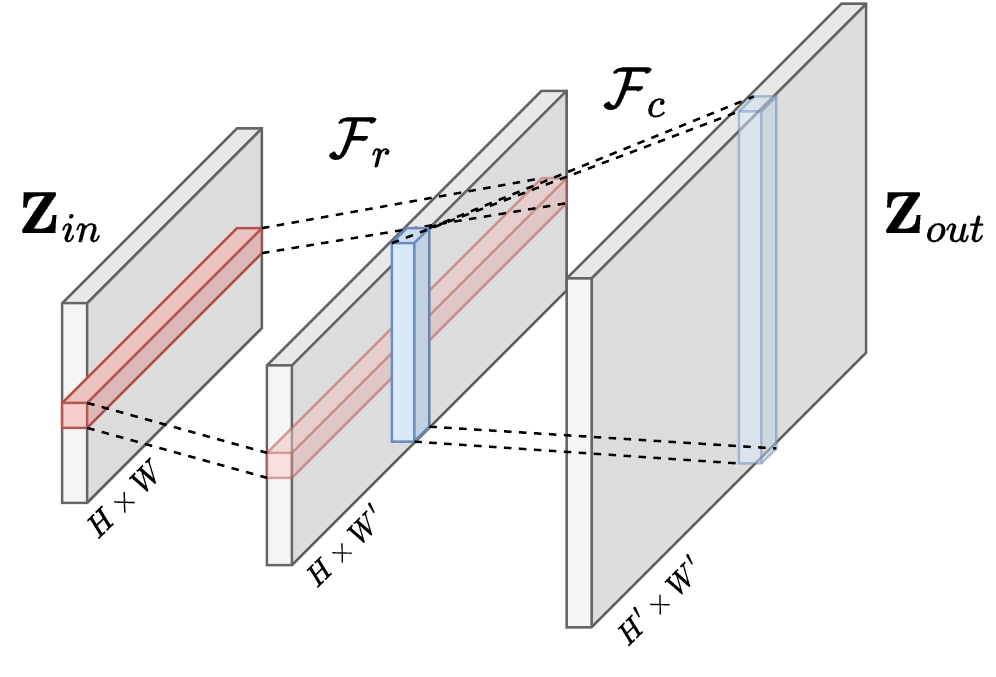}
    \caption{
    The DAFC mechanism concept. Each row of dimension $W$ in $\mathbf{Z}_{in}$, represented by the red color, is transformed by $\mathcal{F}_r$ to a row of dimension $W'$ in the middle matrix, represented by the transparent red color. Next, each column of dimension $H$ in the middle matrix, represented by the blue color, is transformed by $\mathcal{F}_c$ to a column of dimension $H'$ in $\mathbf{Z}_{out}$, represented by the transparent blue color.}\label{fig:DAFC}
\end{figure}

The DAFC block was introduced to process measurements in a form similar to the model in Section~\ref{sec:problem_statement}~\cite{feintuch2022neural}.
Fig.~\ref{fig:DAFC} schematically shows the DAFC mechanism. 

For arbitrary dimensions $D_1,D_2,D_3$, the formulation of a general fully-connected (FC) layer applied to each row in a given matrix $\mathbf{Z}\in\mathbb{R}^{D_1\times D_2}$ can be represented by the transform $\mathcal{F}\left(\cdot\right)$:
\begin{align}\label{eq:matrix_FC_layer}
    &\mathcal{F}:\mathbb{R}^{D_1\times D_2}\rightarrow \mathbb{R}^{D_1\times D_3}\;,\\
    &\mathcal{F}\left(\mathbf{Z}\right) \triangleq h\left(\mathbf{ZW}+\mathbf{1}_{D_1}\mathbf{b}^T\right)\;.\nonumber
\end{align}
This matrix-to-matrix transformation is characterized by the ``learnable'' weight matrix, $\mathbf{W}\in\mathbb{R}^{D_2\times D_3}$, the bias vector, $\mathbf{b}\in\mathbb{R}^{D_3}$, and a scalar element-wise activation function, $h(\cdot)$.

Let $\mathcal{F}_r\left(\cdot\right)$ and $\mathcal{F}_c\left(\cdot\right)$ be two separate, and not necessarily identical instances of $\mathcal{F}\left(\cdot\right)$ from \eqref{eq:matrix_FC_layer}, and $\mathbf{Z}_{in}$ be an arbitrary input matrix.
The DAFC mechanism is formulated by the following operations:
\makeatletter
\renewcommand{\thealgorithm}{}
\renewcommand{\ALG@name}{Dimensional Alternating Fully Connected}
\makeatother

\begin{algorithm}[H]
\caption{}
  \begin{algorithmic}[]
  \Statex \textbullet~\textbf{Input:} $\mathbf{Z}_{in}\in\mathbb{R}^{H\times W}$\\
  \quad\quad\quad\;\;$\mathcal{F}_r:\mathbb{R}^{H\times W}\rightarrow\mathbb{R}^{H\times W'}$\\
  \quad\quad\quad\;\;$\mathcal{F}_c:\mathbb{R}^{W'\times H}\rightarrow\mathbb{R}^{W'\times H'}$
   \begin{enumerate}
   \item Apply a single FC layer to each  row in $\mathbf{Z}_{in}$:
    \begin{align}
        &\mathbf{Z}_r=\mathcal{F}_r\left(\mathbf{Z}_{in}\right)\nonumber
    \end{align}
    \item Apply a single FC layer to each column in $\mathbf{Z}_r$:
    \begin{align}
        &\mathbf{Z}_c=\mathcal{F}_c\left(\mathbf{Z}_r^T\right)\nonumber
    \end{align}
    \item Transpose to keep orientation:
    \begin{align}
        &\mathbf{Z}_{out}=\mathbf{Z}_c^T\nonumber
    \end{align}
    \end{enumerate}
    \Statex \textbullet~\textbf{Output:} $\mathbf{Z}_{out} \triangleq \mathcal{S}\left(\mathbf{Z}\right)\in\mathbb{R}^{H'\times W'}$
      \end{algorithmic}
\end{algorithm}
% This block is repeatedly used in a pipeline structure, such that the input to the first block is the output of the pre-processing flow $\mathbf{Z}_0=\mathcal{P}\left(\mathbf{X}\right)$.
In the following, three DAFC design principles are detailed.

\textbf{1) Structured transformation}
% Conventional DOA estimation considers $K$ snapshots of the length, $L$. The DAFC operation is suitable for processing the same data structure without modifications. 

The input to the first DAFC block is the pre-processed, $\mathbf{Z}_0$, given in~\eqref{eq:Z_0}.
Therefore, the first FC layer, $\mathcal{F}_r$, of the first DAFC block \textcolor{black}{is capable of} extracting spatial-related features from each row in $\mathbf{Z}_0$. 
The second FC layer, $\mathcal{F}_c$, of the first DAFC block, introduces an interaction between transformed rows. 
This implies that a) $\mathcal{F}_r$ \textcolor{black}{can} perform ``spatial-feature'' extraction by transforming \textcolor{black}{each} pre-processed snapshot (each row of $\mathbf{Z}_0$) to a high-dimensional feature space, and b) $\mathcal{F}_c$ \textcolor{black}{can} perform a nonlinear transformation of the extracted features (the columns of $\mathcal{F}_r\left(\mathbf{Z}_0\right)$) from each snapshot.
In this way, the DAFC utilizes both spatial and statistical information. 
% In addition, it can exploit high-order statistics-related features.
Thus, the DAFC mechanism can contribute to estimating the source DOAs and mitigating the interference when incorporated into a NN.

\textbf{2) \textcolor{black}{Dimension reduction of learnable parameters}}

Conventional DOA estimation considers the input data as the collection of measurement vectors (the snapshots $\{\mathbf{x}_k\}$) in a matrix form. 
One straightforward approach to processing the input data using a NN is to reshape it and process it via an FC-based architecture.
In this way, each neuron in the layer's output interacts with every neuron in the input.
On the other hand, the DAFC block transforms the data using a structured transformation, \textcolor{black}{which significantly reduces the number of learnable parameters compared to the straightforward FC-based approach.}
% which is significantly sparser in terms of learnable parameters compared to the straightforward FC-based approach.

This parameter reduction can be observed in the following typical case.
Consider an input matrix $\mathbf{Z}_1\in\mathbb{R}^{D_1\times D_1}$, which is transformed to an output matrix $\mathbf{Z}_2\in\mathbb{R}^{D_2\times D_2}$. The number of learnable parameters in the FC- and the proposed DAFC-based approaches is of the order of $\mathcal{O}\left(D_1^2D_2^2\right)$, and $\mathcal{O}\left(D_1D_2\right)$, respectively. Notice that the DAFC-based transformation complexity grows linearly with the number of learnable parameters compared to the quadratic complexity growth of the straightforward, FC-based approach. 

The contribution of learnable parameters dimension reduction is twofold. 
First, the conventional NN optimization is gradient-based~\cite{Goodfellow-et-al-2016}. Therefore, a significant reduction in the learnable parameter dimension reduces the degrees of freedom in the optimizable parameter space and improves the gradient-based learning algorithm convergence rate. 
Second, reduction in the learnable parameter dimension can be interpreted as increasing the ``inductive bias'' of the NN model~\cite{shalev2014understanding}, which conventionally contributes to the NN statistical efficiency and generalization ability, thus, reducing the NNs tendency to overfit the training data.

\textbf{3) Nonlinearity}

The proposed DAFC considers an additional degree of nonlinearity compared to the straightforward FC-based approach.
A straightforward matrix-to-matrix approach includes an interaction of every neuron in the output matrix with every neuron in the input matrix, followed by an element-wise nonlinear activation function.
On the other hand, the proposed DAFC consists of two degrees of nonlinearity in $\mathcal{F}_r$ and $\mathcal{F}_c$.
Although the weight matrices applied as part of $\mathcal{F}_r$ and $\mathcal{F}_c$ are of lower dimension than the weight matrix used in the straightforward approach, the extra degree of nonlinearity can increase the NN's capacity~\cite{Goodfellow-et-al-2016}. 
Therefore, a NN architecture with the proposed DAFC is capable of learning a more abstract and rich transformation of the input data.

\begin{figure*}[!t]
\centering
\includegraphics[width=0.8\textwidth]{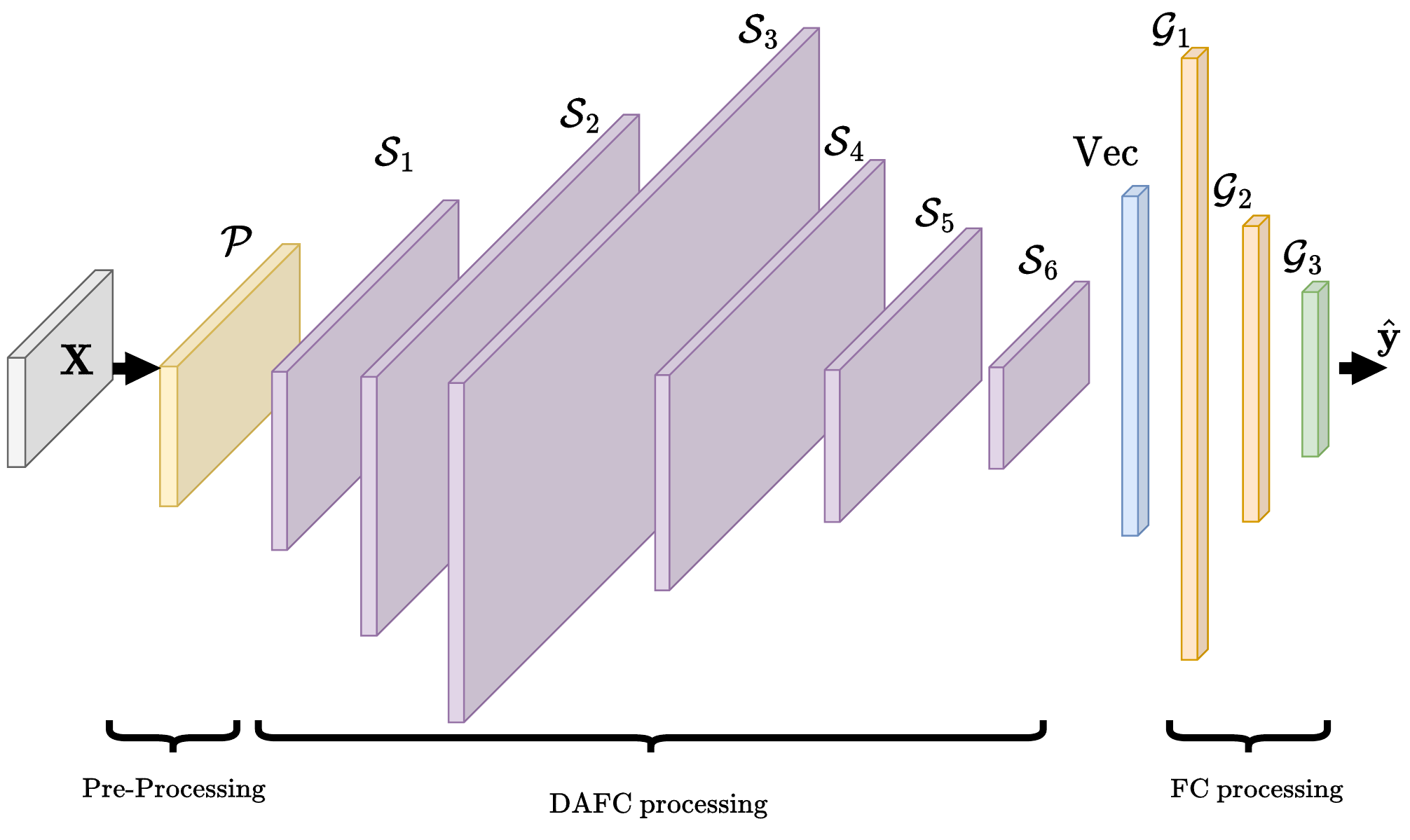}
\caption{Proposed NN architecture. The pre-processing $\mathcal{P}$ is described in Section~\ref{subsec:pre_processing} and appears in yellow. 
The purple matrices denote the concatenation of DAFC blocks, which is detailed in Section~\ref{subsec:dafc}.
The blue vector represents a vectorization of the last DAFC output, and the orange vector stands for FC layers with tanh activation function.
The last green vector is the output of the last FC layer, which consists of the sigmoid activation function and yields the estimated spatial spectrum $\hat{\mathbf{y}}$.
}
\label{fig:NN_arch}
\end{figure*}

\subsection{NN Architecture}\label{subsec:nn_arch}

The continuous DOA space is discretized into a $d$-dimensional grid:  $\bm{\phi}=\begin{bmatrix}\phi_1 & \phi_2 & \cdots & \phi_d \end{bmatrix}^T$.
This implies that the entire field-of-view (FOV) is partitioned into $d$ DOAs, $\{\phi_i\}_{i=1}^{d}$, determined by the selected grid resolution, $\Delta\phi\triangleq\phi_{i+1}-\phi_i$.
The proposed NN is designed to represent a mapping from the input set of snapshots, $\{\mathbf{x}_k\}$ given in~\eqref{eq:meas_model}, into the probability of source present in the DOAs $\{\phi_i\}_{i=1}^{d}$.
The proposed NN architecture is formulated as follows:
\begin{align}\label{eq:NN_arch}
    \mathbf{Z}_0&=\mathcal{P}\left(\mathbf{X}\right)\;,\\
\mathbf{z}_{\text{vec}}&=\text{Vec}\left(\mathcal{S}_6\left(\dots\mathcal{S}_1\left(\mathbf{Z}_0\right)\right)\right)\;,\nonumber\\
\hat{\mathbf{y}}&=\mathcal{G}_3\left(\mathcal{G}_2\left(\mathcal{G}_1\left(\mathbf{z}_{\text{vec}}\right)\right)\right)\;,\nonumber
\end{align}
where $\mathbf{Z}_0$ is the output of the pre-processing procedure, denoted as $\mathcal{P}\left(\cdot\right)$ and detailed in Section~\ref{subsec:pre_processing}, and $\mathbf{X}$ is the input matrix in~\eqref{eq:input_matrix_X}.

In the next stage, six DAFC instances, represented by $\mathcal{S}_1\left(\cdot\right),\dots,\mathcal{S}_6\left(\cdot\right)$, of different dimensions with tanh activation for the row transform ($\mathcal{F}_r$ in Section~\ref{subsec:dafc}) and ReLu activation for the column transform ($\mathcal{F}_c$ in Section~\ref{subsec:dafc}), are used to generate the vectorized signal $\mathbf{z}_{\text{vec}}$.
Our experiments showed that this configuration of row and column activation functions provides the best performance.
At the last stage, the signal, $\mathbf{z}_{\text{vec}}$, is processed by three FC layers, where the first two use tanh activation, and the final (output) layer of equal size to the DOA grid dimension, $d$, use sigmoid activation function to output $\hat{\mathbf{y}}\in\left[0,1\right]^d$. 
Thus, $\{\left[\hat{\mathbf{y}}\right]_i\}_{i=1}^d$ represent the estimated probabilities of a source presence at $\{\phi_i\}_{i=1}^d$.
Table~\ref{tab:NN_arch} and Fig.~\ref{fig:NN_arch} summarize the parameters and architecture of the proposed NN-based approach.

\begin{table}[!t]
\begin{center}
 \begin{tabular}{m{1.5cm} m{2cm} m{1.5cm} m{2cm}}
%  \begin{tabular}{1.0\textwidth,m{4cm} m{1cm} m{1cm}} %{||c c c||}
 \hline\hline
 Operator & Output Dimension & Activation & \# Parameters\\ [1.0ex] 
 \hline\hline
 $\mathcal{P}$ & $K\times 2L$ & - & - \\ [1.0ex]
 \hline
 $\mathcal{S}_1$ & $64\times 256$ & tanh-ReLu & 9,536 \\ [1.0ex]
 \hline
 $\mathcal{S}_2$ & $128\times 512$ & tanh-ReLu & 139,904 \\ [1.0ex]
 \hline
 $\mathcal{S}_3$ & $256\times 1024$ & tanh-ReLu & 558,336 \\ [1.0ex]
 \hline
 $\mathcal{S}_4$ & $64\times 512$ & tanh-ReLu & 541,248 \\ [1.0ex]
 \hline
 $\mathcal{S}_5$ & $16\times 256$ & tanh-ReLu & 132,368 \\ [1.0ex]
 \hline
 $\mathcal{S}_6$ & $4\times 128$ & tanh-ReLu & 32,964 \\ [1.0ex]
 \hline
 $\text{vec}$ & 512 & - & - \\ [1.0ex]
 \hline
 $\mathcal{G}_1$ & 1024 & tanh & 525,312 \\ [1.0ex]
 \hline
 $\mathcal{G}_2$ & 256 & tanh & 262,400 \\ [1.0ex]
 \hline
 $\mathcal{G}_3$ & $d$ & sigmoid & 31,097 \\ [1.0ex]
 \hline
\end{tabular}
\end{center}
\caption{\label{tab:NN_arch} Specification of the proposed NN architecture for $K=16,\;L=16,\;d=121$.
``tanh-ReLu'' activation stands for tanh in $\mathcal{F}_r$ and ReLU in $\mathcal{F}_c$ of each DAFC block.
The number of total learnable parameters is $2,233,165$.}
\end{table}

% \subsection{NN-based DOA Estimation \& Source Enumeration}\label{subsec:NN_DOA_estimation}

The estimated source DOAs are extracted from the spatial spectrum via \textit{peak\_search} and applying $0.5$ threshold:
\begin{align}\label{eq:peak_search_y_hat}
    \{i_1,\dots,i_{\hat{N}}\}&=\textit{peak\_search}\left(\{\left[\hat{\mathbf{y}}\right]_i\}_{i=1}^d\right)\\
    \hat{\mathbf{\Theta}} &=\left\{\phi_{i_n}:\left[\hat{\mathbf{y}}\right]_{i_n}>0.5\right\}_{n=1}^{\hat{N}}\nonumber \; .
\end{align}
Namely, the set of estimated DOAs, $\hat{\mathbf{\Theta}}$, consists of the grid points corresponding to the peaks of $\hat{\mathbf{y}}$ that exceed the $0.5$ threshold.
The number of peaks that exceed this threshold is used for source enumeration, and therefore the proposed NN can be utilized as a source enumeration method as well.

The dimensionality of the hidden layers in the proposed NN architecture expands in the first layers and then reduces.
This trend resembles the NN architecture presented in~\cite{feintuch2022neural} and characterizes both the DAFC-based and FC-based processing stages.
This expansion-reduction structure can be explained by a) the early NN stages need to learn an expressive and meaningful transformation of the input data by mapping it to a higher dimensional representation
and b) the late stages need to extract significant features from the early mappings and are therefore limited in dimensionality.
In addition, the late stages are adjacent to the output vector and therefore need to be of similar dimensions.

\subsection{Loss Function}\label{subsec:nn_training}

The label used for the supervised learning process, $\mathbf{y}\in\{0,1\}^d$, is defined as a sparse binary vector with the value $1$, at the grid points that correspond to the source DOAs, and $0$, otherwise. 
In practice, the DOAs in $\bm{\Theta}$ do not precisely correspond to the grid points. Therefore, for each DOA in $\bm{\Theta}$, the nearest grid point in $\{\phi_i\}_{i=1}^d$ is selected as the representative grid point in the label.
Each training example is determined by the input-label pair, $\left(\mathbf{X},\mathbf{y}\right)$.
Using the NN feed-forward in \eqref{eq:NN_arch}, $\mathbf{X}$ is used to generate the output spatial spectrum, $\hat{\mathbf{y}}$, which is considered as the estimated label.

The loss function, $\mathcal{L}$, is a weighted mean of the binary cross entropy (BCE) loss computed at each grid point:
\begin{align}\label{eq:NN_loss}
    \mathcal{L}\left(\mathbf{y},\hat{\mathbf{y}},t\right)&=\frac{1}{d}\sum_{i=1}^{d}{w_i^{(t)}\text{BCE}\left(\left[\mathbf{y}\right]_i, \left[\hat{\mathbf{y}}\right]_i\right)}\;,\\
    \text{BCE}\left(y,\hat{y}\right)&=-y\log\left(\hat{y}\right) - \left(1-y\right)\log\left(1-\hat{y}\right)\;,\nonumber
    % w_i&=1/p_i\nonumber\;.
\end{align}
where $w_i^{(t)}$ represents the loss weight of the $i$-th grid point at the $t$-th epoch.
The loss value for equally-weighted BCEs evaluated per grid point ($w_i^{(t)}=1$ in \eqref{eq:NN_loss}) does not significantly increase in the case of a large error in source/interference estimated probability due to the sparsity of the label $\mathbf{y}$.
This forces the NN convergence into a sub-optimal solution that is prone to ``miss'' the sources.
Therefore, the loss weights, $\{w_i^{(t)}\}_{i=1}^d$, are introduced to ``focus'' the penalty on source/interference grid points.

The loss weight of the $i$-th grid point, $w_i^{(t)}$, is determined by the presence of source or interference in the corresponding label entry $\left[\mathbf{y}\right]_i$.
This relation is defined using the epoch and label dependent factors $e_{0}^{(t)},e_{1}^{(t)}$, according to:
\begin{align}\label{eq:loss_weights_def}
    w_i^{(t)}&=
    \begin{cases}
         1/e_1^{(t)} ,&\text{if $\phi_i$ contains source or interference}\\
         1/e_0^{(t)} ,&\text{else}
    \end{cases}\;.
\end{align}
For $t=0$, the factor $e_1^{(0)}$ is determined by the fraction of label grid points that contain source or interference out of the total label grid points in the training set, and $e_0^{(0)}$ is the corresponding complement.
For subsequent epochs, the factors are updated according to a predefined schedule, similar to a predefined learning rate schedule.
The loss weights are updated $N_w$ times with spacing of $\Delta t$ epochs during training.
The update values are determined by updating $e_{0}^{(t)},e_{1}^{(t)}$, according to the following decaying rule:
\begin{align}\label{eq:loss_weight_decay}
    e_q^{(t)} &= (1-\beta^{(l)})e_q^{(l\Delta t)} + \beta^{(l)},\;l\Delta t\le t < (l+1)\Delta t\\
    q&=0,1,\;l=1,\dots,N_w,\nonumber
\end{align}
where $l$ is the loss weight update iteration, and $\{\beta^{(l)}\}_{l=1}^{N_w}$ represent the loss weight update factors which uphold, $0\le\beta^{(l)}\le1$.
Note that for $N_w \Delta t\le t$, the weight factor remains $e_i^{(N_w \Delta t)}$ during the rest of the training stage.
Notice that as $\beta^{(l)}\rightarrow1$, the corresponding loss weights will tend to be equally distributed across the grid points, i.e., $e_1^{(t)}\approx e_0^{(t)}$.
In this case, an erroneously estimated probability for source/interference containing grid point is equally weighted to a neither-containing grid point. 
On the other hand, as $\beta^{(l)}\rightarrow0$, the corresponding factors will uphold $e_1^{(t)}\ll e_0^{(t)}$, yielding a significantly larger contribution of source/interference containing grid points to the loss value.
The rule in \eqref{eq:loss_weight_decay} enables a ``transition of focus'' throughout the training. 
That is, during the early epochs $\beta^{(l)}\rightarrow0$, which contributes more weight to the source/interference containing areas in the estimated label $\hat{\mathbf{y}}$ (i.e., the estimated spatial spectrum) to focus the NN to being correct for source/interference.
During the later epochs, $\beta^{(l)}$ is incrementally increased, which relaxes the focus on source/interference from early epochs.
Thus, reducing erroneously estimated sources in areas that do not contain source/interference (i.e., ``false-alarms'').

\section{Performance Evaluation}\label{sec:eval}

This section evaluates the performance of the proposed DAFC-based NN approach and compares it to the conventional approaches, summarized in Subsection~\ref{subsec:benchmarks}.
The data for all considered scenarios are simulated using the measurement model from Section~\ref{sec:problem_statement}.

\subsection{Setup \& Training}
This work considers a uniform linear array (ULA) with half-wavelength-spaced $L$ elements. Each simulated example consists of the input-label pair, $\left(\mathbf{X},\mathbf{y}\right)$, where the input $\mathbf{X}$ is defined in~\eqref{eq:input_matrix_X}, and the label $\mathbf{y}$ is defined in Section~\ref{subsec:nn_training}.
The simulation configurations are detailed in Table~\ref{tab:dataset_config}.
The performance of the proposed approach is evaluated using a \textbf{single NN instance}. Therefore, a single NN model is used for various signal-to-interference ratios (SIRs), signal-to-noise ratios (SNRs), interference-to-noise ratios (INRs), DOAs, interference distribution, and the number of sources for joint DOA estimation and source enumeration. The following definitions for the $m$-th source are used in all experiments: 
\begin{eqnarray}
    \text{INR}&=\frac{\mathbb{E}\left[\left\|\mathbf{c}\right\|^2\right]}{\mathbb{E}\left[\left\|\mathbf{n}\right\|^2\right]}=\sigma_c^2/\sigma_n^2\;,\\
    \text{SNR}_m&=\frac{\mathbb{E}\left[\left\|\mathbf{a}\left(\theta_m\right)s_m\right\|^2\right]}{\mathbb{E}\left[\left\|\mathbf{n}\right\|^2\right]}=\sigma_m^2/\sigma_n^2\;,\\
    \text{SIR}_m&=\frac{\mathbb{E}\left[\left\|\mathbf{a}\left(\theta_m\right)s_m\right\|^2\right]}{\mathbb{E}\left[\left\|\mathbf{c}\right\|^2\right]}=\sigma_m^2/\sigma_c^2\;.
\end{eqnarray}

The NN optimization for all evaluated architectures is performed using the loss function in~\eqref{eq:NN_loss} and Adam optimizer~\cite{kingma2014adam} with a learning rate of $10^{-3}$, and a plateau learning rate scheduler with a decay of $0.905$. 
The set of loss weight update factors, $\{\beta^{(l)}\}_{l=1}^{N_w}$, in \eqref{eq:loss_weight_decay} is chosen as the evenly-spaced logarithmic scale between $10^{-5}$ and $10^{-2}$ with $N_w=6$, that is $\{10^{-5}, 7.25\cdot10^{-5}, 5.25\cdot10^{-4},3.8\cdot10^{-3},2.78\cdot10^{-2},0.2\}$.
The chosen batch size is $512$, the number of epochs is $500$, and early stopping is applied according to the last $200$ epochs.

\begin{table}[ht]
\begin{center}
 \begin{tabular}{m{1.2cm} m{3.5cm} m{2.8cm}}
 \hline\hline
 Notation & Description & Value \\ [1.0ex] 
 \hline\hline
 $M_{max}$ & Maximal number of sources & $4$ \\ [1.0ex]
 \hline
 $L$ & Number of sensors & $16$ \\ [1.0ex]
 \hline
 $K$ & Number of snapshots & $16$ \\ [1.0ex]
 \hline
 $d$ & Angular grid dimension & $121$ \\ [1.0ex]
 \hline
 $\Delta\phi$ & Angular grid resolution & $1\degree$ \\ [1.0ex]
 \hline
 FOV & Field of view & $\left[-60\degree,60\degree\right]$ \\ [1.0ex]
 \hline
 $\sigma_n^2$ & Thermal noise power & $1$ \\ [1.0ex]
 \hline
\end{tabular}
\end{center}
\caption{\label{tab:dataset_config} Simulation Configurations.}
\end{table}

\subsubsection{DOA Estimation Approaches}\label{subsec:benchmarks}
This subsection briefly summarizes the conventional DOA estimation approaches. 
The performance of the proposed approach is compared to the conventional MVDR, CNN, and FC-based NN.
All the NN-based approaches were implemented using a similar number of layers and learnable parameters.
In addition, the FC-based NN and CNN were optimized using the same learning algorithm and configurations.

{\it (a) Conventional Adaptive Beamforming}\label{subsec:mvdr}\\
The MVDR~\cite{capon1969high} estimator is based on adaptive beamforming, and it is  the maximum likelihood estimator in the presence of unknown Gaussian interference~\cite{harmanci2000relationships}.
The MVDR estimates DOAs by a peak search on the MVDR spectrum:
\begin{align}\label{eq:p_mvdr}
    P_{MVDR}\left(\phi\right)=\frac{1}{\mathbf{a}^H\left(\phi\right)\hat{\mathbf{R}}_{x}^{-1}\mathbf{a}\left(\phi\right)}\;,
\end{align}
where $\hat{\mathbf{R}}_x=\frac{1}{K}\sum_{k=1}^{K}{\mathbf{x}_k\mathbf{x}^H_k}$ is the sample covariance matrix estimator. 
Notice that the MVDR spectrum utilizes only second-order statistics of the received signal $\mathbf{x}_k$. For Gaussian-only interference (i.e., $\mathbf{c}_k=0$ in~\eqref{eq:meas_model}), the second-order statistics contain the entire statistical information. However, for non-Gaussian interference, information from higher-order statistics is needed. 

{\it (b) CNN Architecture}\\
We consider a CNN-based DOA estimation approach using a CNN architecture that is similar to the architecture provided in~\cite{papageorgiou2021deep}.
The input to the CNN of dimension $L\times L \times 3$ consists of the real, imaginary, and angle parts of $\hat{\mathbf{R}}_x$.
The CNN architecture consists of $4$ consecutive CNN blocks, such that each block contains a convolutional layer, a batch normalization layer, and a ReLu activation. 
The convolutional layers consist of $\left[128, 256, 256, 128\right]$ filters. 
Kernel sizes of $3\times 3$ for the first block and $2\times 2$ for the following three blocks are used.
Similarly to~\cite{papageorgiou2021deep}, $2\times 2$ strides are used for the first block and $1\times 1$ for the following three blocks.
Next, a flatten layer is used to vectorize the hidden tensor, and $3$ FC layers of dimensions $1024,\;512,\;256$ are used with a ReLu activation and Dropout of $30\%$.
Finally, the output layer is identical to the proposed DAFC-based NN as detailed in Subsection~\ref{subsec:nn_arch}.
The considered loss function is identical to the proposed DAFC-based approach in~\eqref{eq:NN_loss}.
The number of trainable parameters in the considered CNN architecture accounts for $3,315,449$.
Notice that the CNN-based architecture utilizes the information within the sample covariance matrix and, therefore, is limited to second-order statistics only.

{\it (c) FC Architecture}\label{subsubsec:fc_arch}\\
A straightforward implementation of an FC-based architecture, as mentioned in Subsection~\ref{subsec:dafc}, was implemented. 
The data matrix, $\mathbf{X}$, is vectorized, and the real and imaginary parts of the values were concatenated to obtain a $2KL$-dimension input vector.
The selected hidden layers are of sizes: $\left[512, 512, 1024, 1024, 512, 256\right]$ where each hidden layer is followed by a tanh activation function. 
The output layer is identical to the proposed DAFC-based NN approach as detailed in Subsection~\ref{subsec:nn_arch}.
The considered loss function is~\eqref{eq:NN_loss}, and the number of trainable parameters in the FC-based NN accounts for $2,787,449$.
Notice that the FC-based NN architecture utilizes all the measurements by interacting with all samples in the input data.
However, this processing is not specifically tailored to the structure of information within the measurements.
On the other hand, the proposed DAFC-based NN utilizes the information structure to process the input data.
Therefore, for the considered DOA estimation problem, the ``inductive bias''~\cite{shalev2014understanding} for this approach is improper and can result in under-fitted NN architecture.

\subsubsection{Performance Evaluation Metrics}\label{subsec:eval_metrics}
This subsection discusses the criteria for the performance evaluation of the proposed DOA estimation approach.
In this work, similarly to~\cite{papageorgiou2021deep}, the DOA estimation accuracy of a set of sources is evaluated by the Hausdorff distance between sets.
% In this work, the DOA estimation accuracy of a set of sources is evaluated by the Hausdorff distance between sets~\cite{papageorgiou2021deep}.
The Hausdorff distance, $d_H$ between the sets, $\mathcal{A}$, and $\mathcal{B}$, is defined as:
\begin{align}\label{eq:d_H}
    d_H\left(\mathcal{A},\mathcal{B}\right)&= \max\left\{d\left(\mathcal{A},\mathcal{B}\right), d\left(\mathcal{B},\mathcal{A}\right)\right\}\;,\\
    d\left(\mathcal{A},\mathcal{B}\right)&=\sup\left\{\inf\left\{\left|\alpha-\beta\right| : \beta\in\mathcal{B} \right\}: \alpha\in\mathcal{A} \right\}\;.\nonumber
\end{align}
Notice that $d\left(\mathcal{A},\mathcal{B}\right)\ne d\left(\mathcal{B},\mathcal{A}\right)$. 
Let $\mathbf{\Theta}=\{\theta_m\}_{m=1}^M$ and $\hat{\mathbf{\Theta}}=\{\hat{\theta}_m\}_{m=1}^{\hat{M}}$ be the sets of true and estimated DOAs, respectively.
The estimation error is obtained by evaluating the Hausdorff distance, $d_H(\mathbf{\Theta}, \hat{\mathbf{\Theta}})$.
We define the root mean squared distance (RMSD) for an arbitrary set of $N$ examples (e.g., test set), $\left\{\mathbf{X}^{\left(n\right)},\mathbf{y}^{\left(n\right)}\right\}_{n=1}^N$, with the corresponding true and estimated DOAs, $\left\{\mathbf{\Theta}^{\left(n\right)},\hat{\mathbf{\Theta}}^{\left(n\right)}\right\}_{n=1}^{N}$ as:
\begin{align}\label{eq:RMSD}
    \text{RMSD}&\triangleq\sqrt{\frac{1}{N}\sum_{n=1}^{N}{d_H^2\left(\mathbf{\Theta}^{\left(n\right)}, \hat{\mathbf{\Theta}}^{\left(n\right)}\right)}}\;.
\end{align}

Angular resolution is one of the key criteria for DOA estimation performance. 
The probability of resolution is commonly used as a performance evaluation metric for angular resolution. 
In the considered problem, resolution between two sources and between source and interference are used for performance evaluation. 
For an arbitrary example with $M$ sources, the resolution event $A_{res}$ is defined as:
\begin{align}\label{eq:A_res}
    A_{res}\left(\mathbf{\Theta},\hat{\mathbf{\Theta}}\right)&\triangleq
    \begin{cases}
        1,& \bigcap_{m=1}^{M}{\xi_m \le 2\degree} \text{ and } |\hat{\mathbf{\Theta}}|\ge M\\
        0,& \text{else}
    \end{cases}\;,\\
    \xi_m&\triangleq\min_{\hat{\theta}\in\hat{\mathbf{\Theta}}}|\theta_m - \hat{\theta}|,\;m=1,\dots,M\;.\nonumber
\end{align}
For example, a scene with $M$ sources is considered successfully resolved if for each true DOA a) there exists a close-enough estimated DOA, $\hat{\theta}\in\hat{\mathbf{\Theta}}$, that is at most $2\degree$ apart, and b) there exists at least $M$ DOA estimations.
According to \eqref{eq:RMSD}, the probability of resolution can be defined as:
\begin{align}\label{eq:P_res}
    P_{res}&=\frac{1}{N}\sum_{n=1}^{N}{A_{res}\left(\mathbf{\Theta}^{\left(n\right)},\hat{\mathbf{\Theta}}^{\left(n\right)}\right)}\;.
\end{align}

\subsubsection{Data Sets}
This subsection describes the structure and formation of \textit{Training} \& \textit{Test sets}.

{\it (a) Training Set}\\
The considered training set contains $N_{train}=10,000$ examples re-generated at each epoch. 
For each example, i.e. an input-label pair $\left(\mathbf{X},\mathbf{y}\right)$, the number of DOA sources, $M$, is generated from uniform and $i.i.d.$ distribution, $\{1,\dots,M_{max}\}$. 
The training set contains $10\%$ of interference-free examples and $90\%$ of interference-containing.
Out of the interference-containing examples, $90\%$ generated such that the source DOAs, $\{\theta_m\}_{m=1}^M$, and the interference's DOA, $\theta_c$, are distributed uniformly over the simulated FOV. 
The remaining $10\%$ are generated such that $\theta_c$ is distributed uniformly over the FOV, and the source DOAs, $\{\theta_m\}_{m=1}^M$, are distributed uniformly over the interval $\left[\theta_c-8\degree,\theta_c+8\degree\right]$.
This data set formation enables to ``focus'' the NN training on the challenging scenarios where the source and interference DOAs are closely spaced.
The generalization capabilities of the proposed NN to variations in interference statistics are achieved via the interference angular spread parameter, $\rho$, from the uniform distribution, $\text{U}\left(\left[0.7,0.95\right]\right)$, and the interference spikiness parameter, $\nu$, from the uniform distribution, $\text{U}\left(\left[0.1,1.5\right]\right)$. 
The $\text{INR}$ for each interference-containing example and $\{\text{SIR}_m\}_{m=1}^{M}$ or $\{\text{SNR}_m\}_{m=1}^{M}$ are drawn independently according to Table~\ref{tab:train_params}.

{\it (b) Test Set}\\
The test set consists of $N_{test}=20,000$ examples.
The results are obtained by averaging the evaluated performance over $50$ independent test set realizations. 
Considering the low-snapshot support regime, the number of snapshots is set to $K=16$, except for experiment (c) in \ref{subsubsec:2source_K}. 
Considering heavy-tailed interference, the spikiness parameter is set to $\nu=0.2$.
The INR is set to $\text{INR}=5\;dB$, and the interference angular spread parameter is set to $\rho=0.9$.
% Considering scenarios with the heavy-tailed distributed interference, the $\text{INR}$ and $\nu$ parameters were set to $\text{INR}=5\;dB$, and $\nu=0.2$, respectively.
The signal amplitude was set to be identical for all sources, $\sigma_1=\dots=\sigma_m$, except for experiment (b) in \ref{subsubsec:2source_unequal}.

\begin{table}[ht]
\begin{center}
 \begin{tabular}{m{1cm} m{3cm} m{3.5cm}}
 \hline\hline
 Notation & Description & Value \\ [1.0ex] 
 \hline\hline
 $\rho$ & Interference angular spread parameter& $\sim\text{U}\left(\left[0.7,0.95\right]\right)$ \\ [1.0ex]
 \hline
 $\nu$ & Interference spikiness parameter & $\sim\text{U}\left(\left[0.1,1.5\right]\right)$ \\ [1.0ex]
 \hline
 $\text{INR}$ & $\text{INR}$ & $\sim\text{U}\left([0,10]\right)\;[dB]$ \\ [1.0ex]
 \hline
 $\text{SIR}_m$ & SIR of $m$-th source & $\sim\text{U}\left([-10,10]\right)\;[dB]$ \\ [1.0ex]
 \hline
 $\text{SNR}_m$ & $\text{SNR}$ of $m$-th source & $\sim\text{U}\left([-10,10]\right)\;[dB]$ \\ [1.0ex]
 \hline
\end{tabular}
\end{center}
\caption{\label{tab:train_params} Training set parameters. $\text{SNR}_m$ distribution applies to interference-free examples.}
\end{table}

\begin{figure*}[!t]
    \begin{subfigure}{0.49\textwidth}
    \caption{}
    \includegraphics[width=\linewidth]{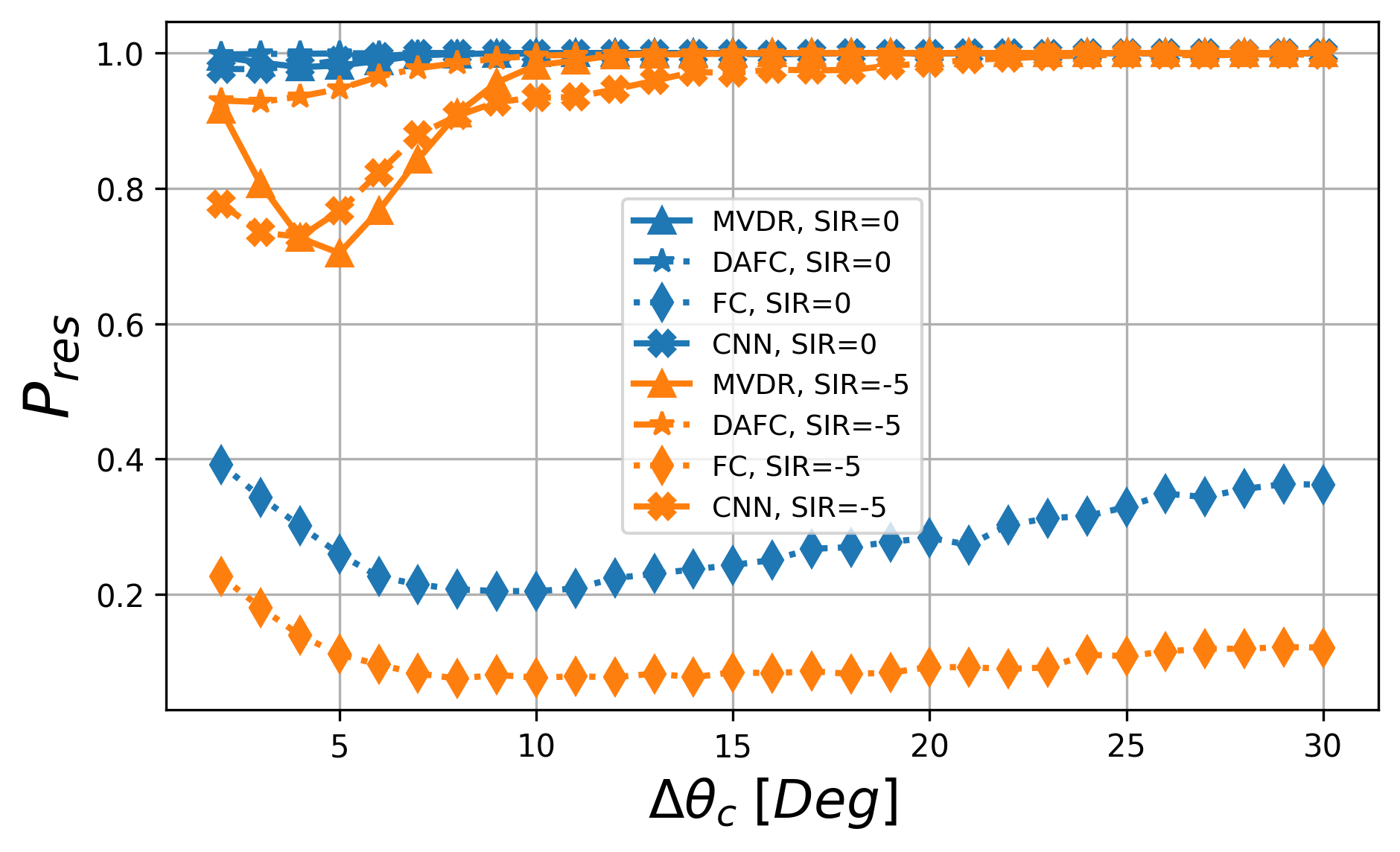}
    \label{subfig:interf_1target_P_res}
  \end{subfigure}%
  \hspace*{\fill}   % maximize separation between the subfigures
  \begin{subfigure}{0.49\textwidth}
    \caption{}
    \includegraphics[width=\linewidth]{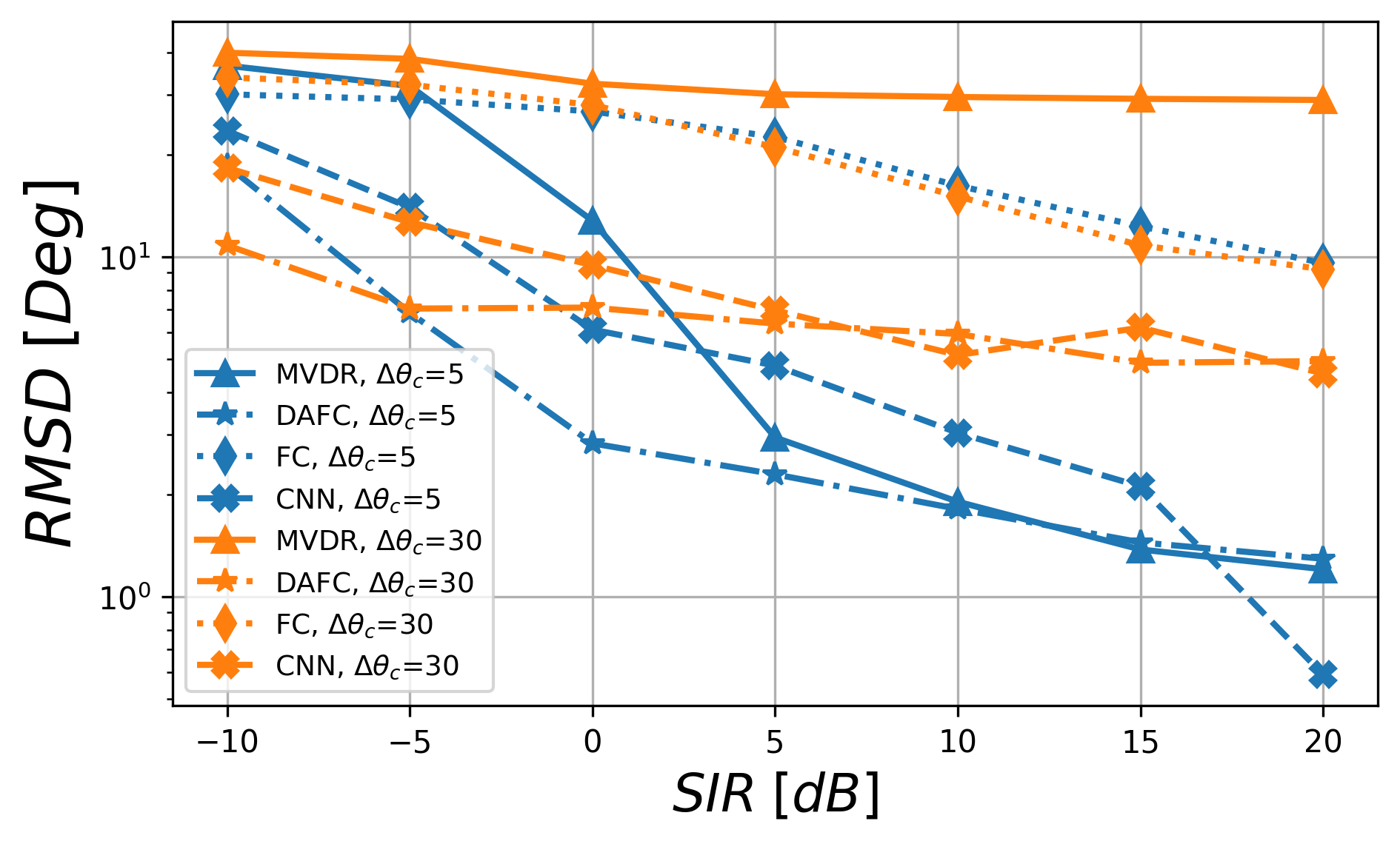}
    \label{subfig:interf_1target_P_RMSD}
  \end{subfigure}%
\caption{Scenario with a single source at $\theta_1=0.55\degree$ and interference located at $\theta_c=\theta_1+\Delta\theta_c$. (a) probability of resolution and (b) RMSD.} 
\label{fig:interf_1_source}
\end{figure*}

\subsection{Experiments}
\subsubsection{Single Source Within Interference}\label{subsubsec:eval_1_source}
In this scenario, the ability to resolve a single source from interference is evaluated.
Let $M=1$ with $\theta_1=0.55\degree$, and $\theta_c=\theta_1+\Delta\theta_c$ such that $\Delta\theta_c$ is the angular separation between the single source and interference.
The $0.55\degree$ offset is considered to impose a realistic off-grid condition.
Fig.~\ref{fig:interf_1_source} shows the RMSD and probability of resolution for all evaluated approaches.

Fig.~\ref{subfig:interf_1target_P_res} shows that the FC-based NN approach does not manage to resolve the single source from the interference for all evaluated angular separations.
This result supports the under-fitting limitation of the FC-based NN approach for the DOA estimation, which can be explained by the architecture that processes the input data as-is, without any structured transformation or model-based pre-processing. The MVDR and CNN performance in terms of resolution are similar since both of them rely only on second-order statistics, which is sufficient in scenarios with widely separated sources and interference. \textcolor{black}{It can be seen in Fig.~\ref{subfig:interf_1target_P_res} that for $2\degree<\Delta\theta_c<5\degree$ the MVDR shows a drop, which is a result of the source's proximity to the interference. When the angular distance between the interference and the source reduces below $5^\circ$, the source and interference peaks coincide, and thus the probability  of resolution is reduced. When the angular distance further decreases, the probability that the united peak falls within the resolution window increases, resulting in higher $P_{res}$.}
Fig.~\ref{subfig:interf_1target_P_res} shows that the proposed DAFC-based NN approach outperforms all other considered approaches in low angular separation scenarios. 
% This can be explained by the fact that the DAFC uses the high-order statistics needed for the resolution of closely spaced sources and interference. 
\textcolor{black}{This observation serves as an evidence of the capability of the DAFC to utilize the information in the input data that is needed for resolution between closely spaced sources and interference.}

Fig.~\ref{subfig:interf_1target_P_RMSD} shows the RMSD of all considered DOA estimation approaches. The proposed DAFC-based NN approach outperforms the other tested approaches in low SIR.
At high SIR and small angular separation, $\Delta\theta_c=5\degree$, the interference is negligible with respect to the strong source signal, and therefore, the DAFC-based, CNN, and MVDR approaches obtain similar performance.
For large angular separation, $\Delta\theta_c=30\degree$, the source and the interference are sufficiently separated, and therefore, DOA estimation errors are mainly induced by the interference DOA, $\theta_c$.
The MVDR spectrum contains a peak at $\theta_c=30.55\degree$, and therefore, MVDR's $RMSD=30\degree$ is approximately constant. 
The NNs are trained to output a $0$-probability for the interference. Therefore, the NN-based approaches: FC, CNN, and DAFC achieve a smaller DOA estimation error.
The DAFC-based NN and CNN utilize structured transformations, which better fit the input data, and therefore, they outperform the FC-based NN approach in terms of RMSD.

\begin{figure*}[ht]
    
\begin{subfigure}{0.49\textwidth}
    \caption{}
    \includegraphics[width=\linewidth]{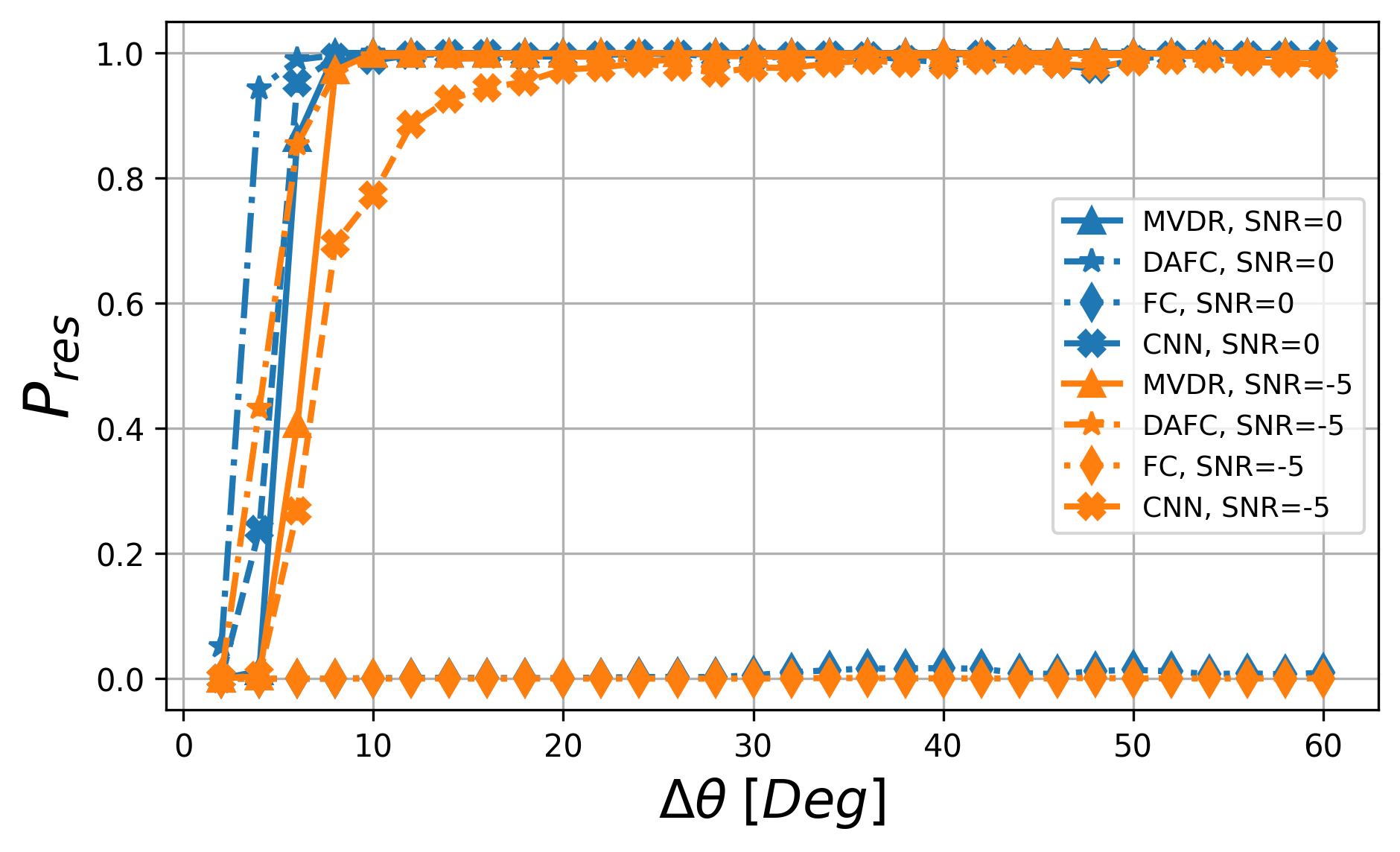}
    \label{subfig:interf_2target_P_res_WGN}
\end{subfigure}%
  \hspace*{\fill}   % maximize separation between the subfigures
\begin{subfigure}{0.49\textwidth}
    \caption{}
    \includegraphics[width=\linewidth]{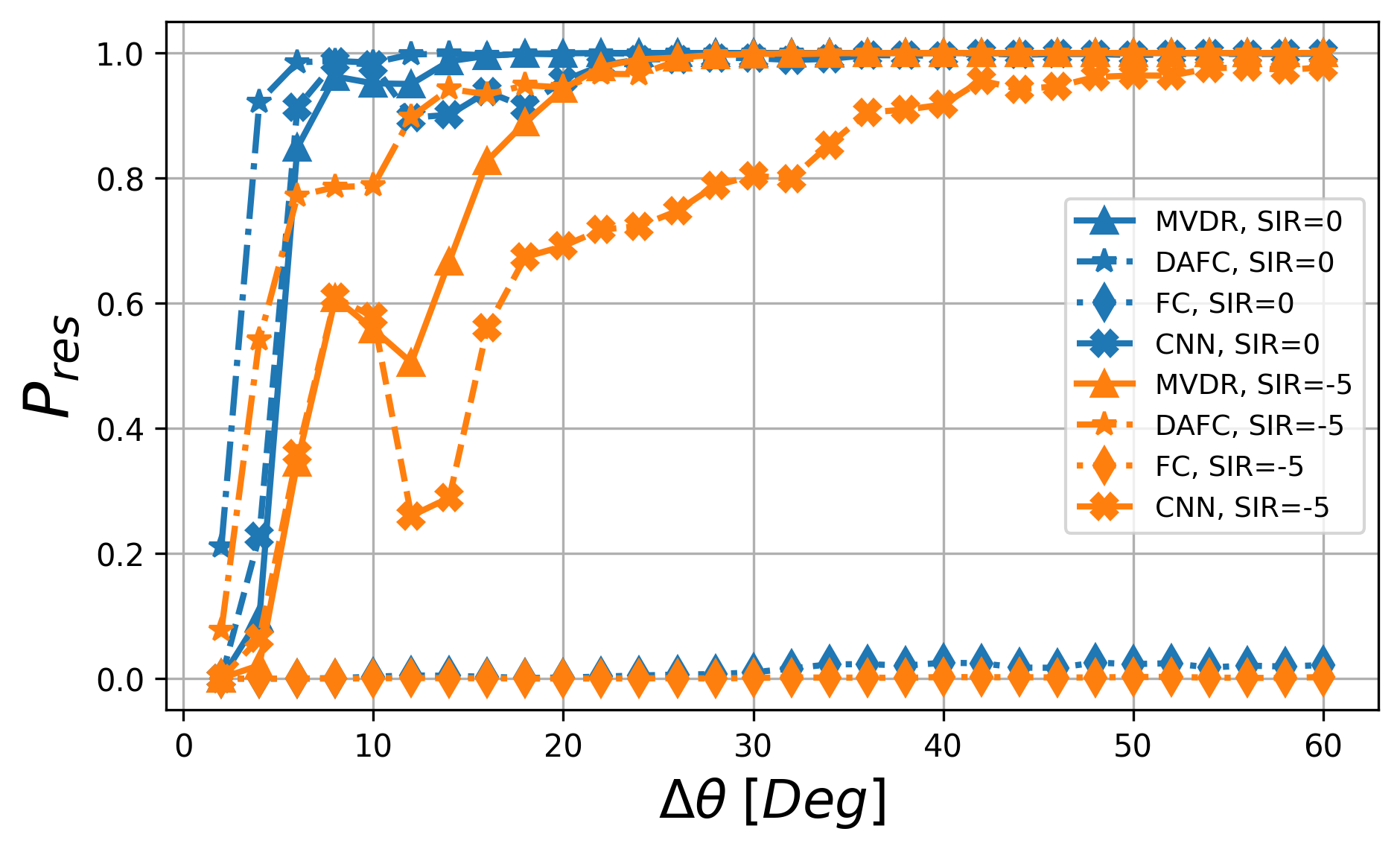}
    \label{subfig:interf_2target_P_res}
\end{subfigure}%

\caption{
Probability of resolution for two sources located at $\theta_{1,2}=\theta_c\pm\Delta\theta/2$, and interference located at $\theta_c=0.55\degree$. (a) AWGN-only scenario and (b) interference-containing scenario.} 
\label{fig:interf_2_source}
\end{figure*}

\begin{figure}[ht]
\centering
  \begin{subfigure}{0.25\textwidth}
    \caption{FC}
    \includegraphics[width=\linewidth]{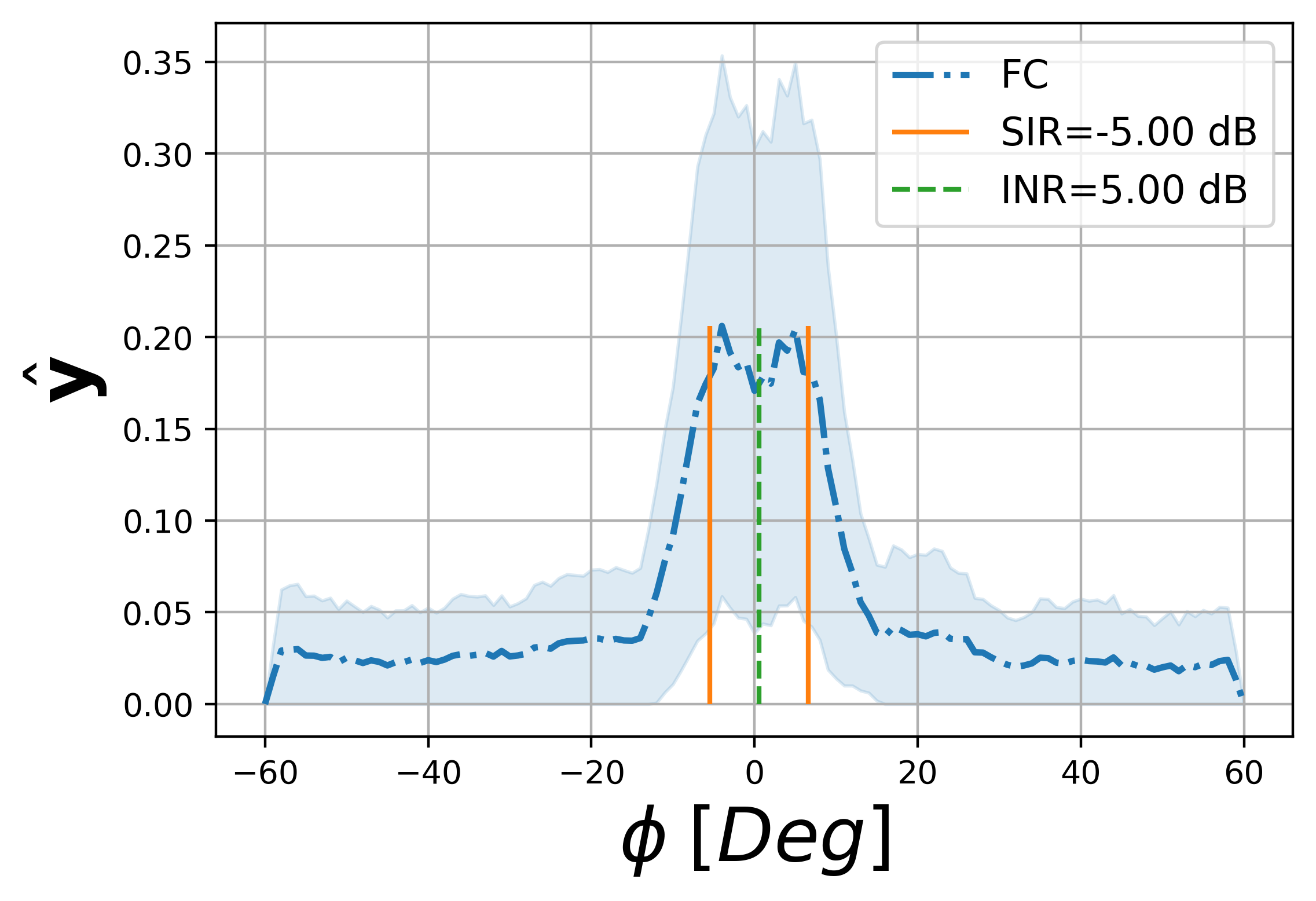}
  \label{subfig:fc_spectrum_delta_12_sir_-5}
  \end{subfigure}%
%   \vspace*{\fill}   % maximize separation between the subfigures
  \begin{subfigure}{0.25\textwidth}
    \caption{MVDR}
    \includegraphics[width=\linewidth]{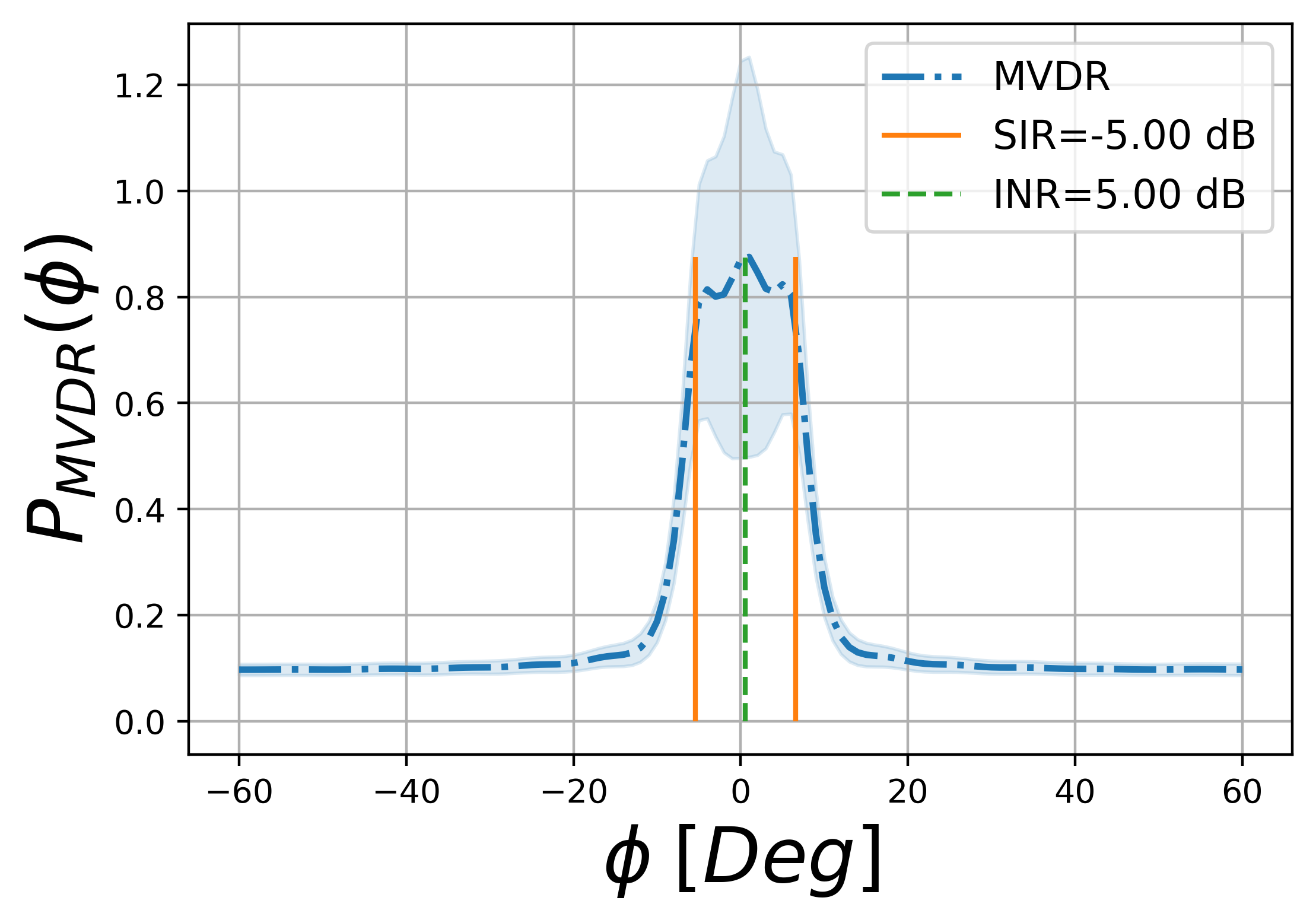}
  \label{subfig:mvdr_spectrum_delta_12_sir_-5}
  \end{subfigure}%
  
  \vspace*{\fill}   % maximize separation between the subfigures
  \begin{subfigure}{0.25\textwidth}
    \caption{CNN}
    \includegraphics[width=\linewidth]{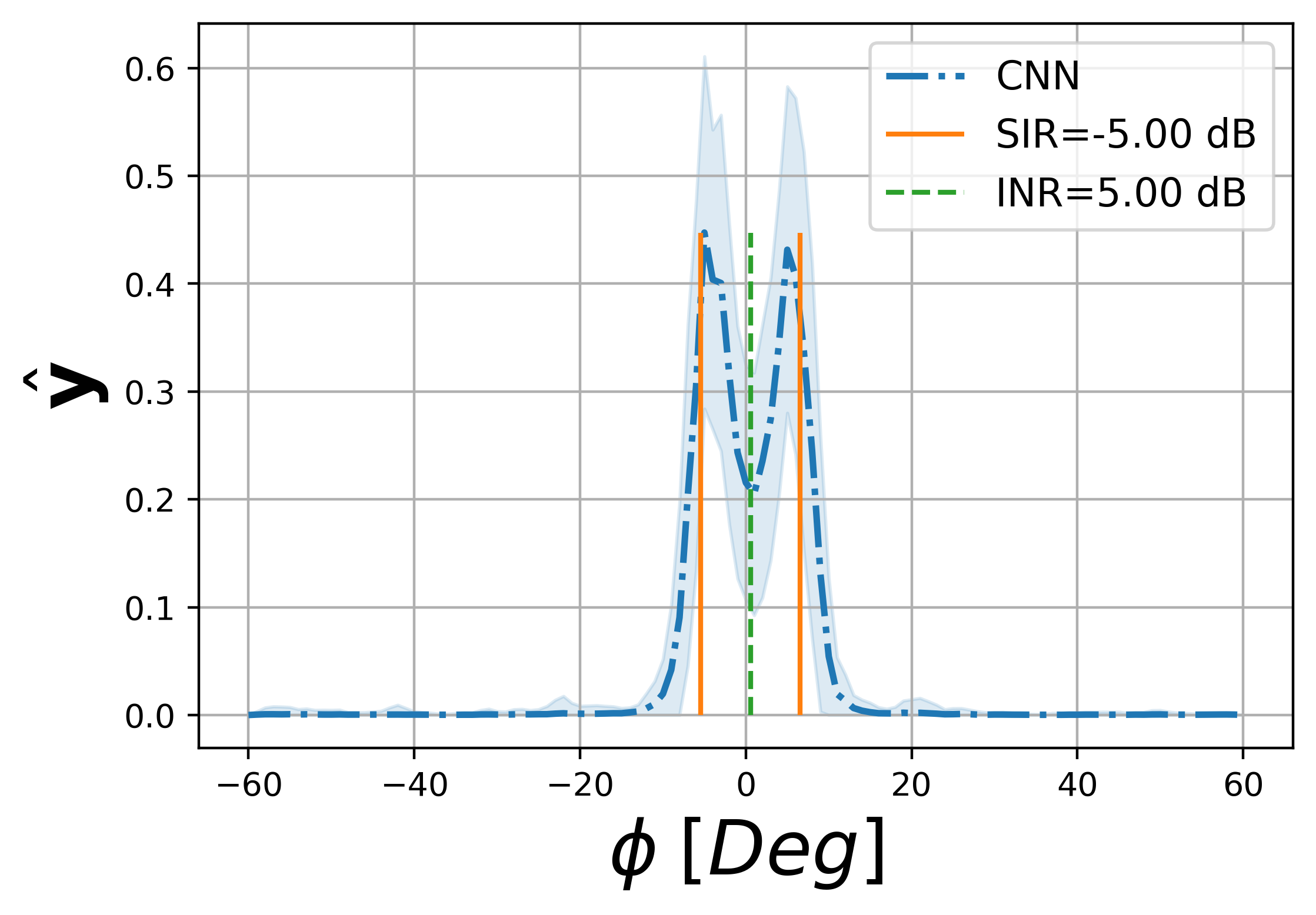}
  \label{subfig:cnn_spectrum_delta_12_sir_-5}
  \end{subfigure}%
%   \vspace*{\fill}   % maximize separation between the subfigures
  \begin{subfigure}{0.25\textwidth}
    \caption{DAFC}
    \includegraphics[width=\linewidth]{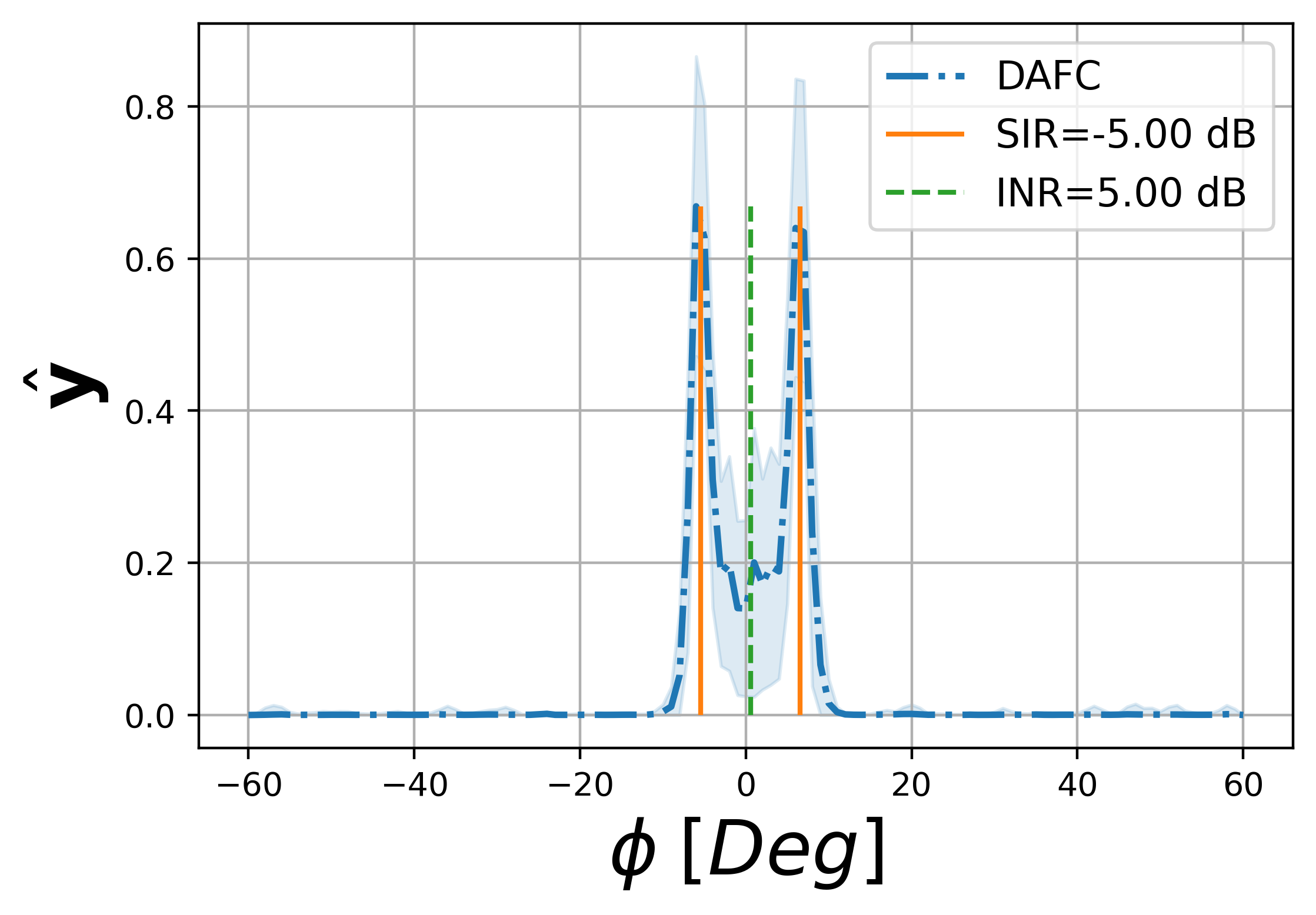}
  \label{subfig:dafc_spectrum_delta_12_sir_-5}
  \end{subfigure}%
  
\caption{
Spatial spectrum, two sources with $\text{SIR}=-5\;dB$ located at $\theta_{1,2}=\theta_c\pm\Delta\theta/2$ with $\Delta\theta=12\degree$ and $\theta_c=0.55\degree$.
The dashed blue lines represent the mean spatial spectrum, and the color fill represents the standard deviation around the mean obtained from $2,000$ i.i.d. examples.
The solid vertical orange lines represent the true source DOAs, and the dashed vertical green line represents the interference DOA.
} 
\label{fig:interf_two_sources_Spectrum}

\end{figure}

\subsubsection{Resolving Two Sources from Interference}\label{subsec:two_source_res}
This subsection evaluates the performance of the tested DOA estimation approaches in scenarios with two sources within AWGN and interference.

{\it (a) Resolution of Equal-Strength Sources}\\
In the following experiment, the resolution between two equal-power sources, $M=2$, with $\theta_1=-\frac{\Delta\theta}{2}+0.55\degree$, and $\theta_2=\frac{\Delta\theta}{2}+0.55\degree$, is evaluated. 
The off-grid additional $0.55\degree$ offset to the $\Delta\theta$ angular separation between the sources represents the practical scenario.
The interference at $\theta_c=0.55\degree$ influences the two sources similarly.
Fig.~\ref{fig:interf_2_source} shows the probability of resolution of the tested approaches in scenarios with (a) the AWGN only and (b) spatially-colored interference.

The FC-based NN approach does not resolve the two targets in both evaluated scenarios.
Subplot (a) in Fig.~\ref{fig:interf_2_source} shows that the proposed DAFC-based NN approach outperforms the MVDR and the CNN at low-SNR and small angular separation scenarios due to its generalization ability to spatially-white interference.
Subplot (b) in Fig.~\ref{fig:interf_2_source} shows that at low SIR of $\text{SIR}=-5\;dB$, the performances of MVDR and CNN significantly degrade compared to the proposed DAFC-based NN approach.
Comparing subplots in Fig.~\ref{fig:interf_2_source}, notice that at $\text{SIR}=-5\;dB$, the MVDR fails to resolve the sources with angular separation $\Delta\theta<20\degree$ due to the presence of the heavy-tailed spatially-colored interference in the proximity of the sources. 
However, the proposed DAFC-based NN approach mitigates this interference and resolves the sources, and hence, outperforms other tested approaches at both $\text{SIR}=0 \;dB$ and $\text{SIR}=-5 \;dB$.

Subplot (b) in Fig.~\ref{fig:interf_2_source} shows the non-monotonic trend of CNN and MVDR performance at $4\degree<\Delta\theta<18\degree$ and $\text{SIR}=-5\;dB${\color{red}, which was observed also in Fig.~\ref{subfig:interf_1target_P_res}}.
For $4\degree<\Delta\theta<8\degree$, the sources are closer to the peak of the interference's lobe and are, therefore, less mitigated by it.
As $\Delta\theta$ increases in the range $8\degree<\Delta\theta<12\degree$, the sources reach DOAs which are in the proximity of the interference lobe's ``nulls'' which explains the reduction in resolution, and as $\Delta\theta$ further increases, $16\degree<\Delta\theta$, the sources are sufficiently separated from the interference such that the resolution increases.
As a result, MVDR and CNN-based approaches that use second-order statistics only can not resolve the sources in the vicinity of a stronger interference.

Fig.~\ref{fig:interf_two_sources_Spectrum} shows the average spatial spectrum of all tested approaches for $\Delta\theta=12\degree$ and $\text{SIR}=-5\;dB$.
The average spatial spectrum of the FC-based NN approach does not show two prominent peaks, which results in its poor probability of resolution in Fig.~\ref{fig:interf_2_source}.
The MVDR ``bell-shaped'' spatial spectrum does not contain the two prominent peaks at $\theta_{1,2}$ since the interference ``masks'' the two sources.
The CNN and proposed DAFC-based NN approaches show two peaks at the average spatial spectrum.
The peaks at the CNN's average spatial spectrum are lower, resulting in a low-resolution probability.
The average spatial spectrum of the proposed DAFC-based NN approach contains two high peaks, resulting in a superior probability of resolution in Fig.~\ref{fig:interf_2_source}.

\begin{figure}[ht]
    \centering
    \includegraphics[width=0.49\textwidth]{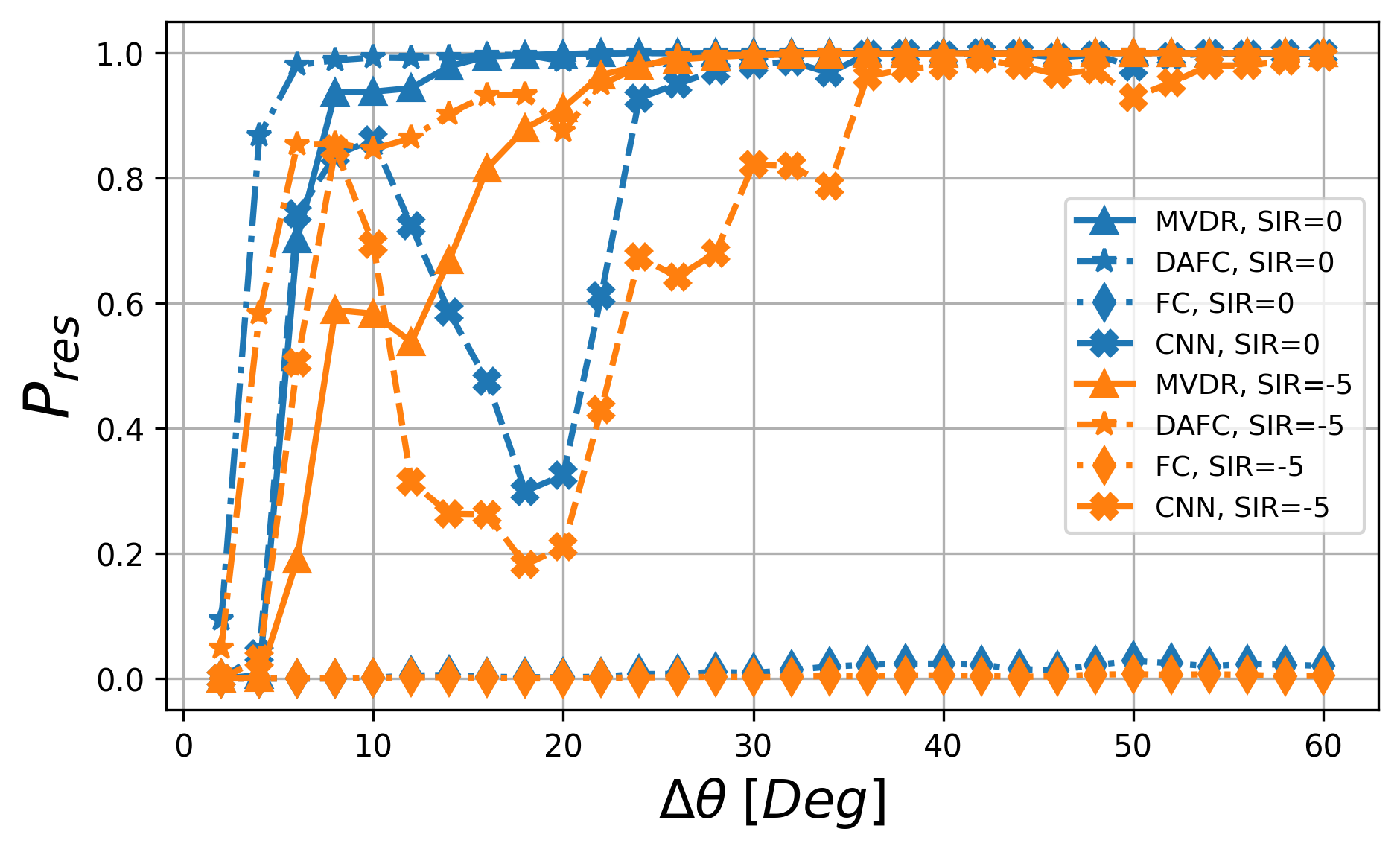}
    \caption{Probability of resolution for two sources located at $\theta_{1,2}=\theta_c\pm\Delta\theta/2$, and interference located at $\theta_c=0.55\degree$. 
    The SIR in the legend represents the SIR of the first source, $\text{SIR}_1$. The SIR of the second source is set to $\text{SIR}_2=\text{SIR}_1+10\;dB$. 
    } 
    \label{fig:interf_2_source_delta_sir}
\end{figure}

{\it (b) Resolution of Unequal-Power Sources}\label{subsubsec:2source_unequal}\\
Fig.~\ref{fig:interf_2_source_delta_sir} shows the probability of resolution in a scenario with two sources, $M=2$, at $\theta_{1}=-\Delta\theta/2+0.55\degree$, and $\theta_{2}=+\Delta\theta/2+0.55\degree$ with interference located between the sources at $\theta_c=0.55\degree$.
The signal strength of the second source is set to $\text{SIR}_1=\text{SIR}_2+10\;dB$.
Comparing Fig.~\ref{fig:interf_2_source_delta_sir} to Fig.~\ref{subfig:interf_2target_P_res}, the competing methods show similar trends, except the degradation of the CNN's probability of resolution for the $\text{SIR}=0\;dB$ case. On the other hand, the proposed DAFC-based NN approach outperforms other tested approaches in terms of the probability of resolution.
Therefore, Fig.~\ref{fig:interf_2_source_delta_sir} demonstrates the generalization ability of the proposed DAFC-based NN approach to a variance between source strengths.
\textcolor{black}{The reason for the drop of $P_{res}$ of CNN and MVDR in $10\degree<\Delta\theta<20\degree$ is similar to the same phenomenon that appears in Figs.~\ref{subfig:interf_1target_P_res} and \ref{subfig:interf_2target_P_res}, which was explained in Subsection~\ref{subsec:two_source_res}(a).}

\begin{figure}[ht]
    \centering
    \includegraphics[width=0.49\textwidth]{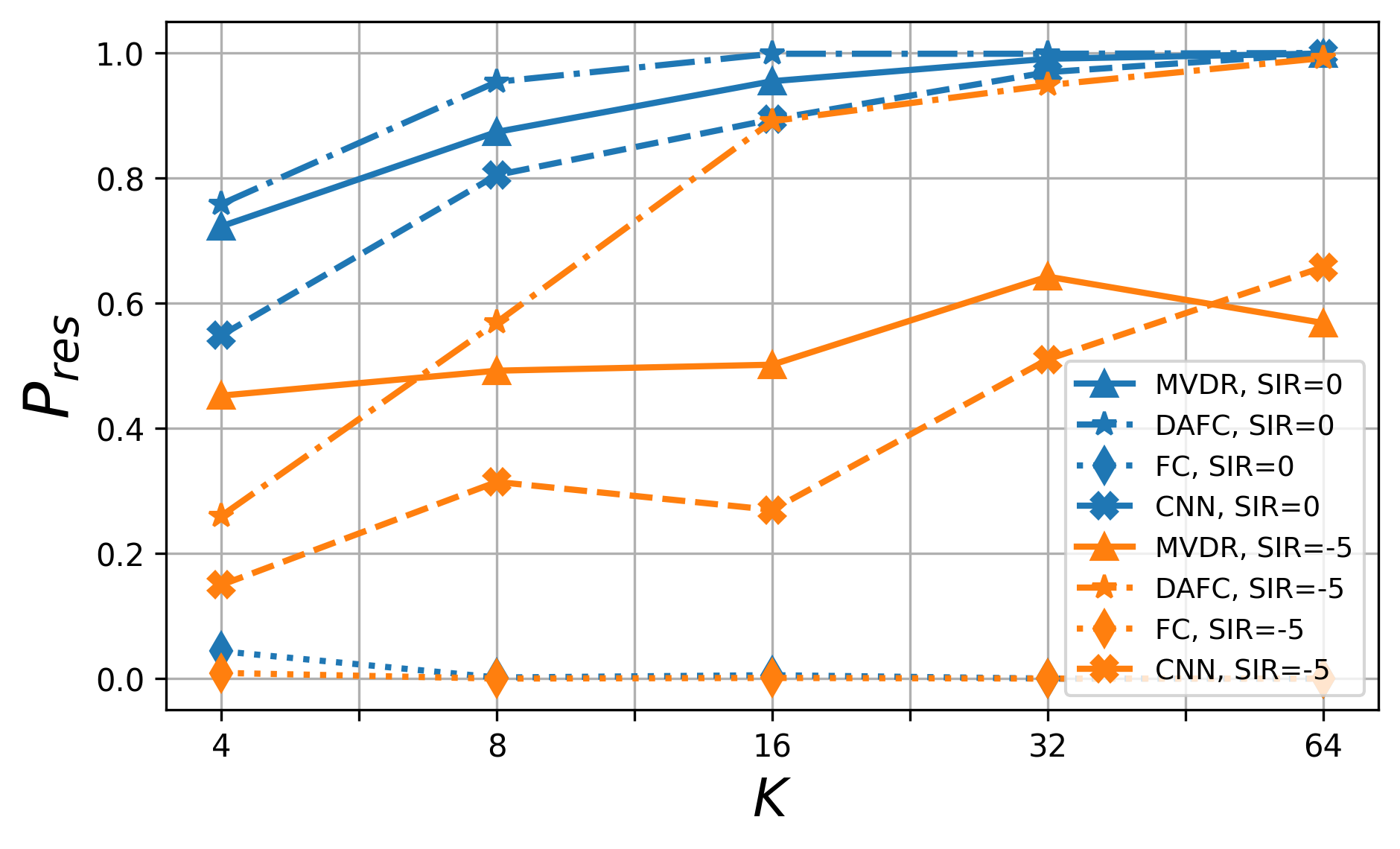}
    \caption{
    Probability of resolution for two sources located at $\theta_{1,2}=\theta_c\pm\Delta\theta/2$ with $\Delta\theta=12\degree$, and interference located at $\theta_c=0.55\degree$, as a function of the number of snapshots, $K$.
    } 
    \label{fig:K_snapshots}
\end{figure}

{\it (c) Effect of the Number of Snapshots on the Resolution}\label{subsubsec:2source_K}\\
This experiment investigates the influence of the number of snapshots, $K$, on the ability to resolve two proximate sources from heavy-tailed spatially-colored interference.
The equal-strength resolution scenario is repeated using $K=4,\;8,\;16,\;32,\;64$ with different instances of NN training for each $K$ value.
Fig.~\ref{fig:K_snapshots} shows the probability of resolution for two equal-strength sources at $\theta_{1,2}=\theta_c\pm\Delta\theta/2$ for $\Delta\theta=12\degree$ and $\theta_c=0.55\degree$.

The FC-based NN approach fails to resolve the two sources.
For $\text{SIR}=0\;dB$, the MVDR, CNN, and DAFC-based NN approaches achieve a monotonic increasing probability of resolution with increasing $K$. The proposed DAFC-based NN approach slightly outperforms other tested approaches.
At low SIR of $\text{SIR}=-5\;dB$, the proposed DAFC-based NN approach significantly outperforms the other tested approaches.
This can be explained by the fact that increasing $K$ increases the probability for outliers to be present in the input data matrix, $\mathbf{X}$. 
Therefore, the estimated autocorrelation matrix, $\hat{\mathbf{R}}_{x}$, is more likely to be biased by the interference-related outliers, which results in interference ``masking'' the sources. 
The proposed DAFC-based NN approach is immune to these outliers and successfully exploits the information from the additional snapshots to improve the probability of resolution.

Figs. \ref{fig:interf_2_source}, \ref{fig:interf_two_sources_Spectrum}, \ref{fig:interf_2_source_delta_sir}, and \ref{fig:K_snapshots} show the ability of the proposed DAFC-based NN approach to utilize the information structure of the input data by performing the domain-fitted transformation in order to provide superior resolution ability in the case of proximate heavy-tailed spatially-colored interference, low SIR and small sample size.

\subsubsection{Multiple Source Localization}

\begin{figure}[ht]
    \centering
    \includegraphics[width=0.49\textwidth]{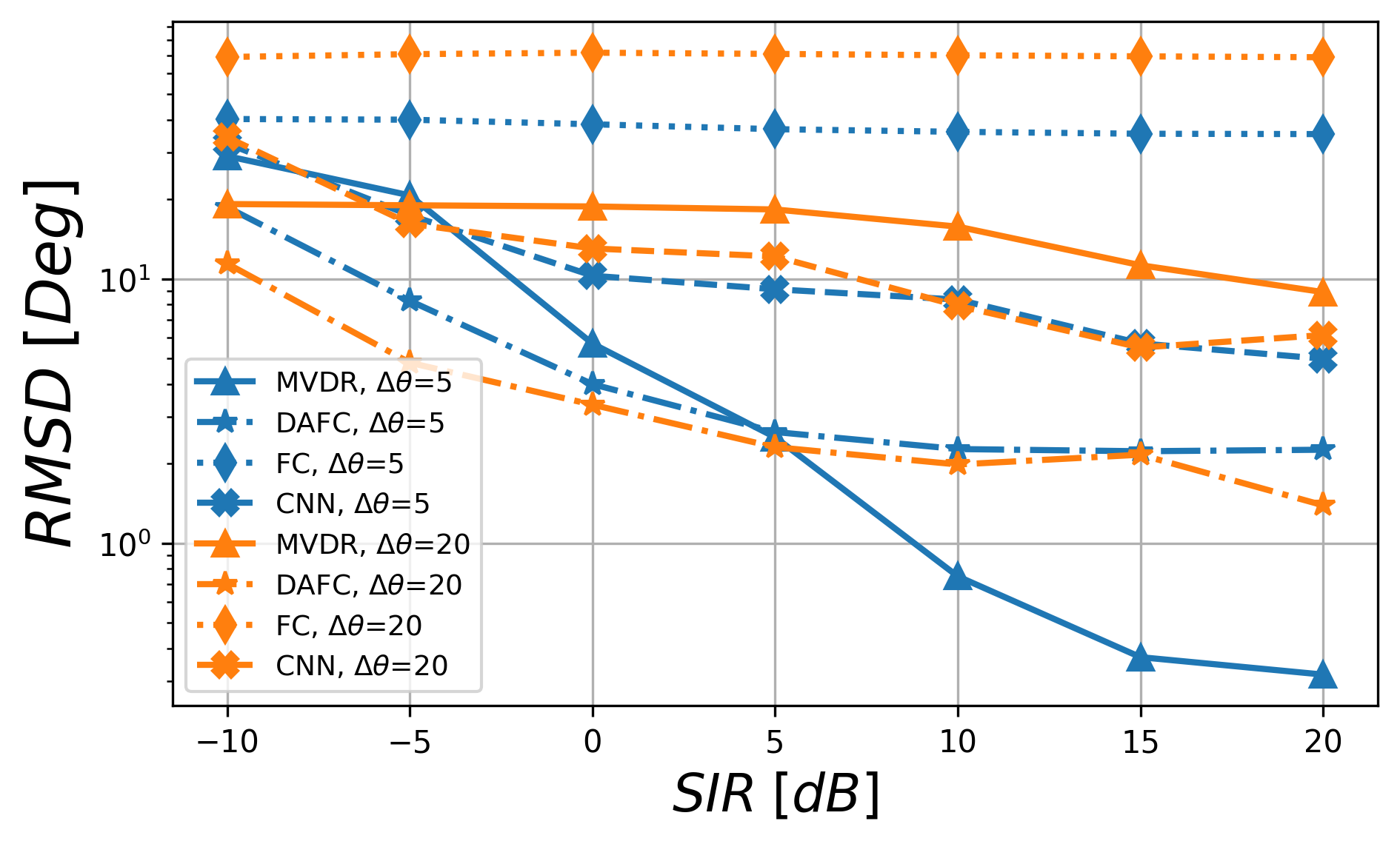}
    \caption{RMSD in scenarios with $M=4$ sources located at $\{\theta_1,\theta_2,\theta_3,\theta_4\}=\theta_c + \{-2\Delta\theta, -\Delta\theta, \Delta\theta, 2\Delta\theta\}$, where \textcolor{black}{$\theta_c\sim\text{U}\left(-15^\circ,15^\circ \right)$}.} 
    \label{fig:multi_sources}
\end{figure}

\begin{figure}[ht]
\centering
  \begin{subfigure}{0.25\textwidth}
    \caption{FC}
    \includegraphics[width=\linewidth]{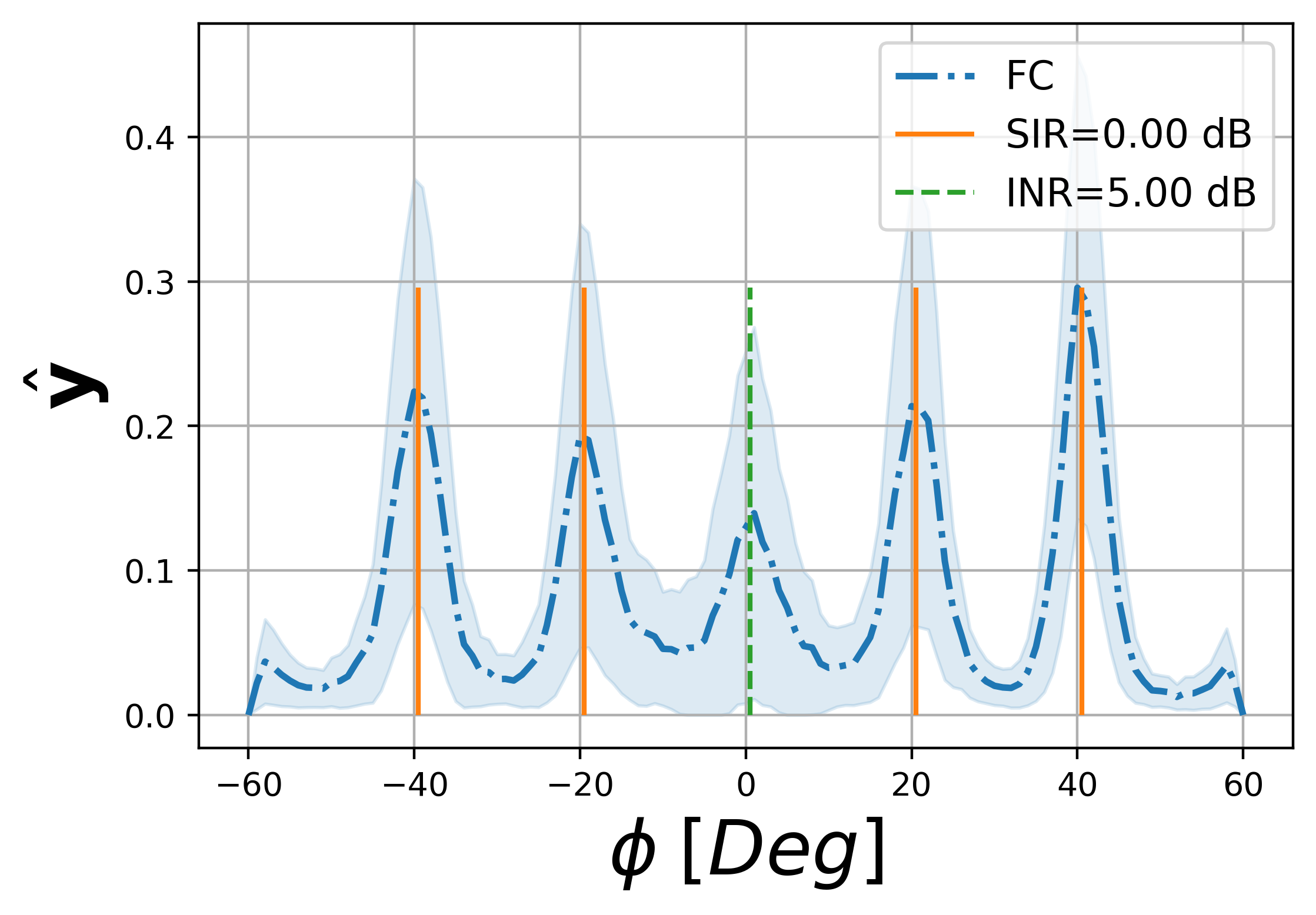}
  \label{subfig:multi_source_fc}
  \end{subfigure}%
  \hspace*{\fill}   % maximize separation between the subfigures
  \begin{subfigure}{0.25\textwidth}
    \caption{MVDR}
    \includegraphics[width=\linewidth]{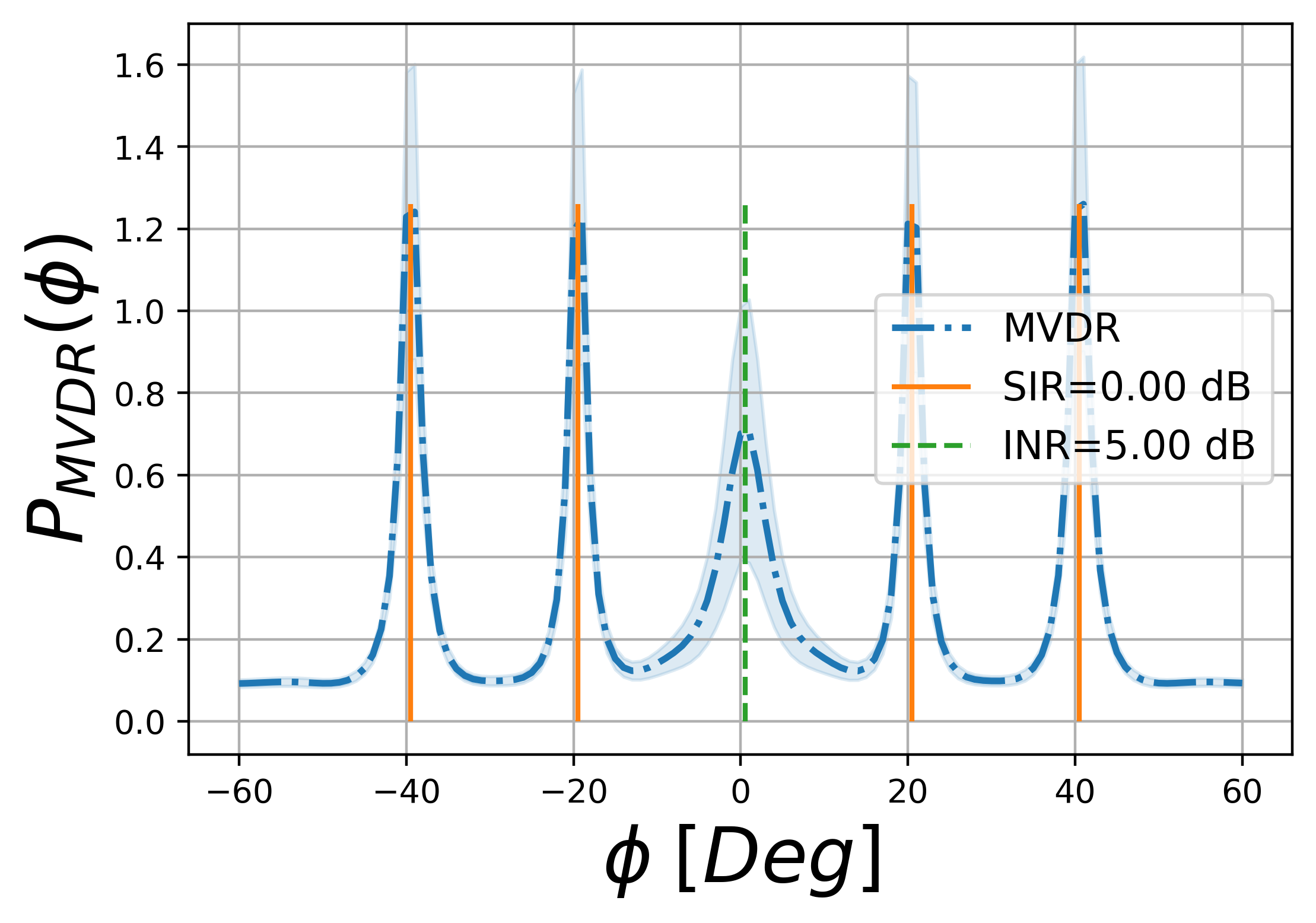}
  \label{subfig:multi_source_mvdr}
  \end{subfigure}%
  
  \vspace*{\fill}   % maximize separation between the subfigures
  \begin{subfigure}{0.25\textwidth}
    \caption{CNN}
    \includegraphics[width=\linewidth]{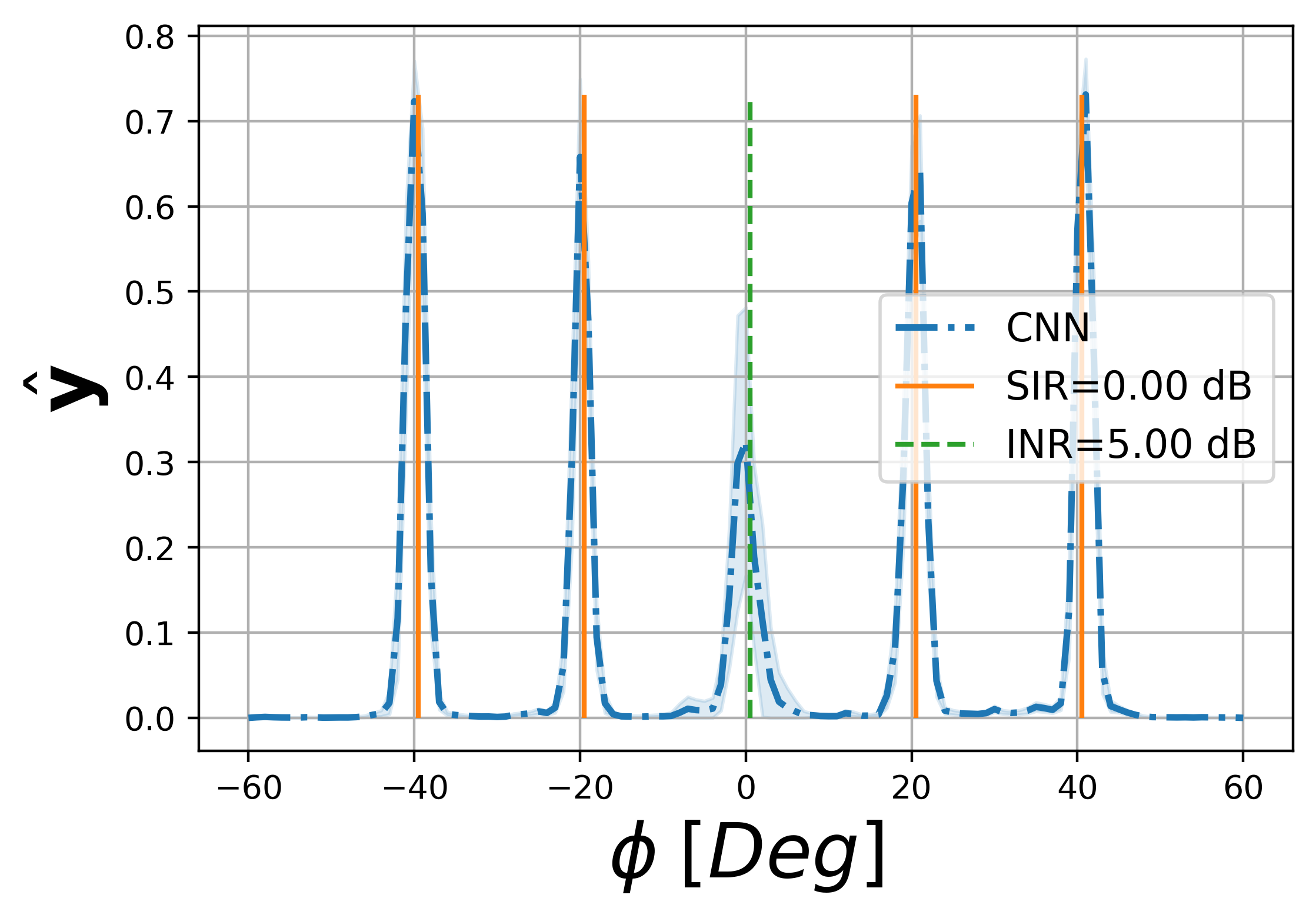}
  \label{subfig:multi_source_cnn}
  \end{subfigure}%
  \hspace*{\fill}   % maximize separation between the subfigures
  \begin{subfigure}{0.25\textwidth}
    \caption{DAFC}
    \includegraphics[width=\linewidth]{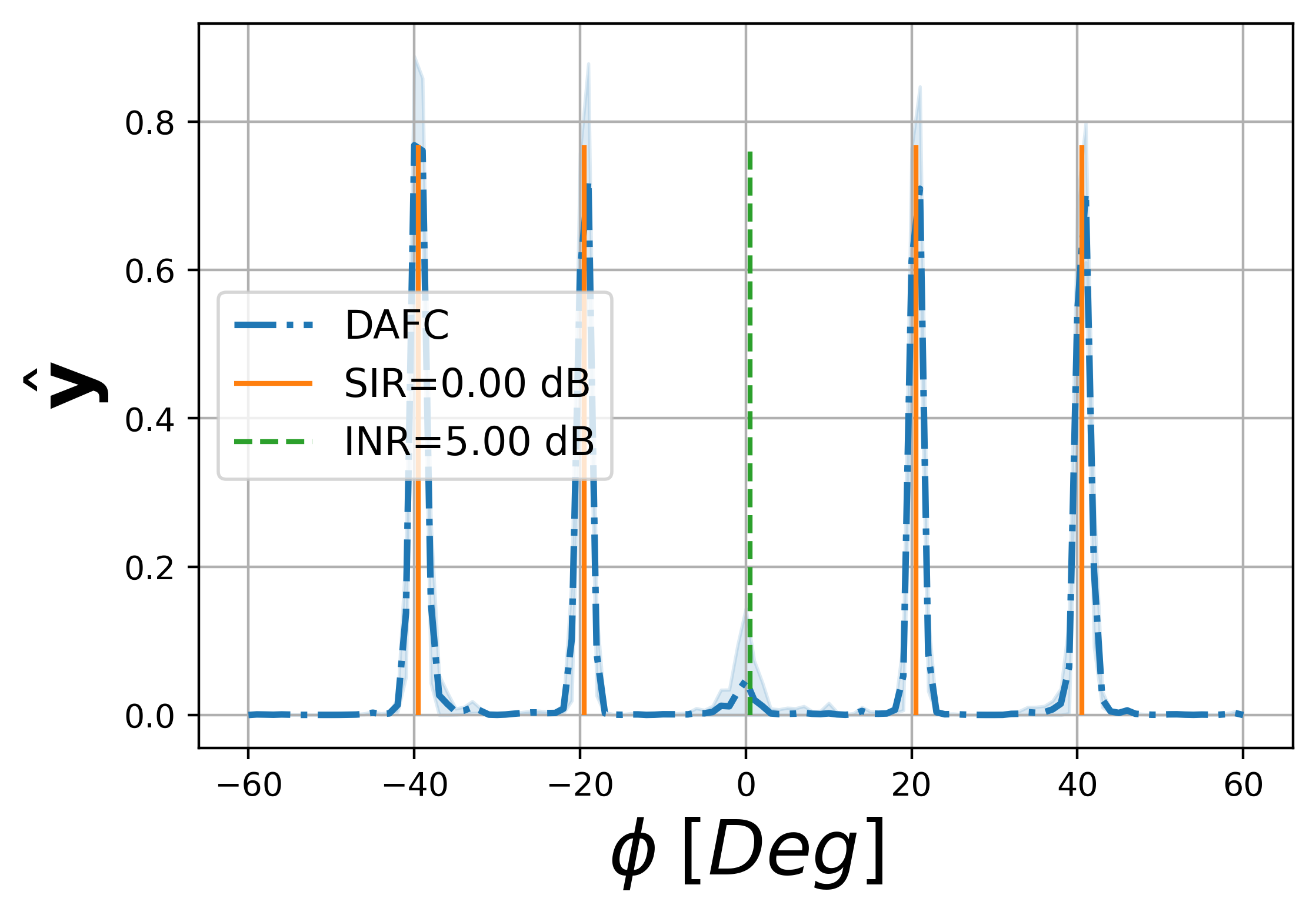}
  \label{subfig:multi_source_dafc}
  \end{subfigure}%
  
\caption{
Spatial spectrum, four sources with $\text{SIR}=0\;dB$ located at $\{\theta_1,\theta_2,\theta_3,\theta_4\}=\theta_c + \{-2\Delta\theta, -\Delta\theta, \Delta\theta, 2\Delta\theta\}$, where $\theta_c=0.51\degree$ and $\Delta\theta=20\degree$. 
The dashed blue lines represent the mean spatial spectrum, and the color fill represents the standard deviation around the mean obtained from $2,000$ i.i.d. examples.
The solid vertical orange lines represent the true source DOAs, and the dashed vertical green line represents the interference DOA.} 
\label{fig:multi_source_spectrum}
\end{figure}

The performances of the tested DOA estimation approaches are evaluated and compared in a multi-source scenario. Four sources, ($M=4$) were simulated with angular separation, $\Delta\theta$: $\{\theta_1,\theta_2,\theta_3,\theta_4\}=\theta_c + \{-2\Delta\theta, -\Delta\theta, \Delta\theta, 2\Delta\theta\}$, where \textcolor{black}{$\theta_c\sim\text{U}\left(-15^\circ,15^\circ\right)$} represents a realistic condition \textcolor{black}{for source and interference DOAs}.
The RMSD of evaluated methods is depicted in Fig.~\ref{fig:multi_sources}.
The proposed DAFC-based NN approach outperforms the other tested approaches at low SIR ($\text{SIR}<0\;dB$) for large and small angular separations.
For high SIR and low angular separation, $\Delta\theta=5\degree$, the MVDR achieves the lowest RMSD.
The reason is that, for this case, the interference is negligible with respect to the lobe of the strong source in the MVDR's spectrum. 
However, at high angular separation, $\Delta\theta=20\degree$, the proposed DAFC-based NN approach significantly outperforms the other tested approaches.
This is explained by Fig.~\ref{fig:multi_source_spectrum}, which shows the spectrum of the tested DOA estimation approaches. 
Notice that the proposed DAFC-based NN mitigates interference, while the spectra of other tested approaches contain high peaks at the interference DOA, $\theta_c$. 
These peaks increase the Hausdorff distance in \eqref{eq:d_H}, increasing the RMSD of other tested approaches in Fig.~\ref{fig:multi_sources}.

\begin{figure}[ht]
    \centering
  \begin{subfigure}{0.35\textwidth}
    \caption{MDL}
    \includegraphics[width=\linewidth]{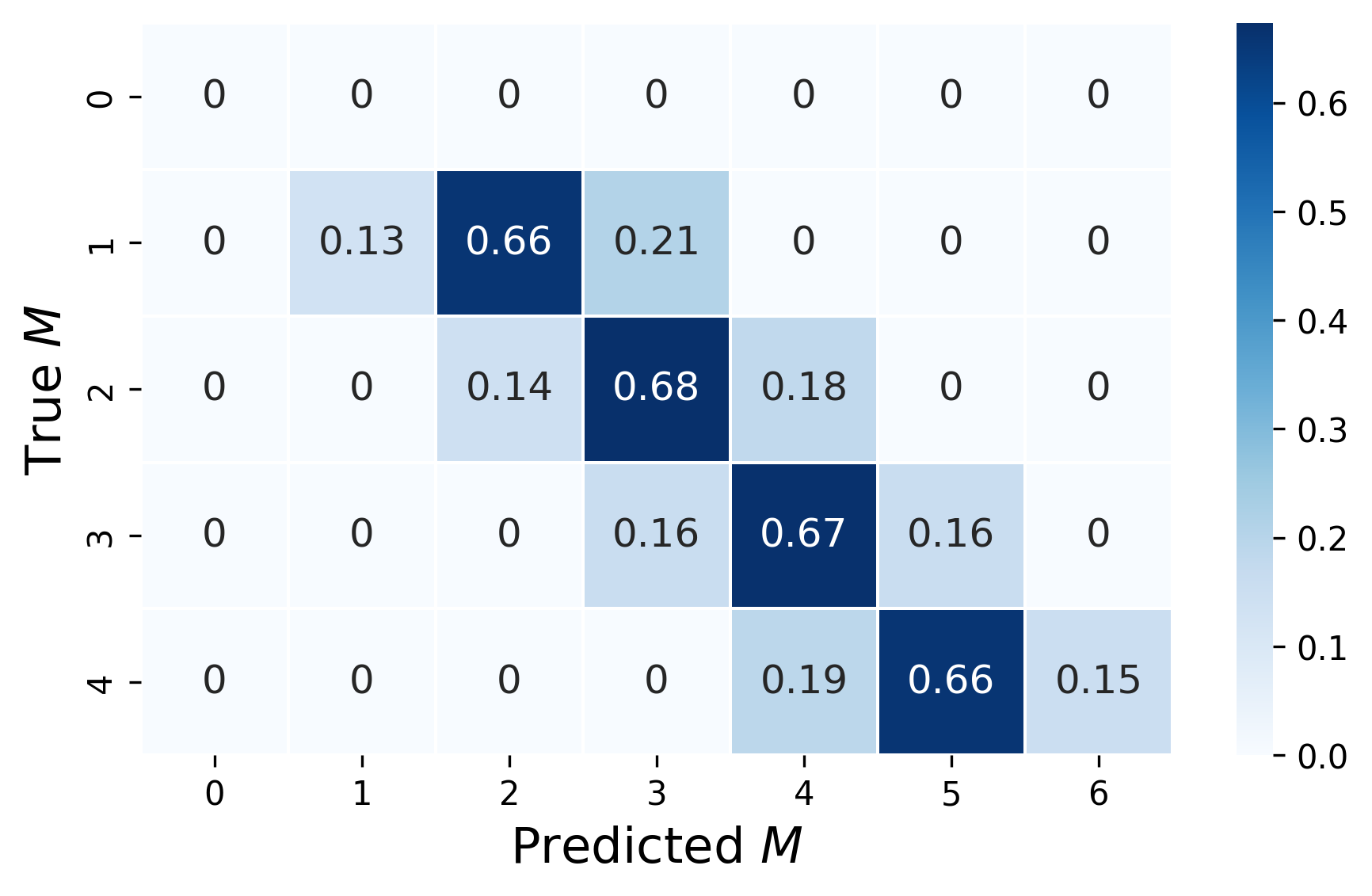}
    \label{subfig:source_enum_conf_mat_MDL}
  \end{subfigure}%
  \vspace*{\fill}   % maximize separation between the subfigures
  \begin{subfigure}{0.35\textwidth}
    \caption{AIC}
    \includegraphics[width=\linewidth]{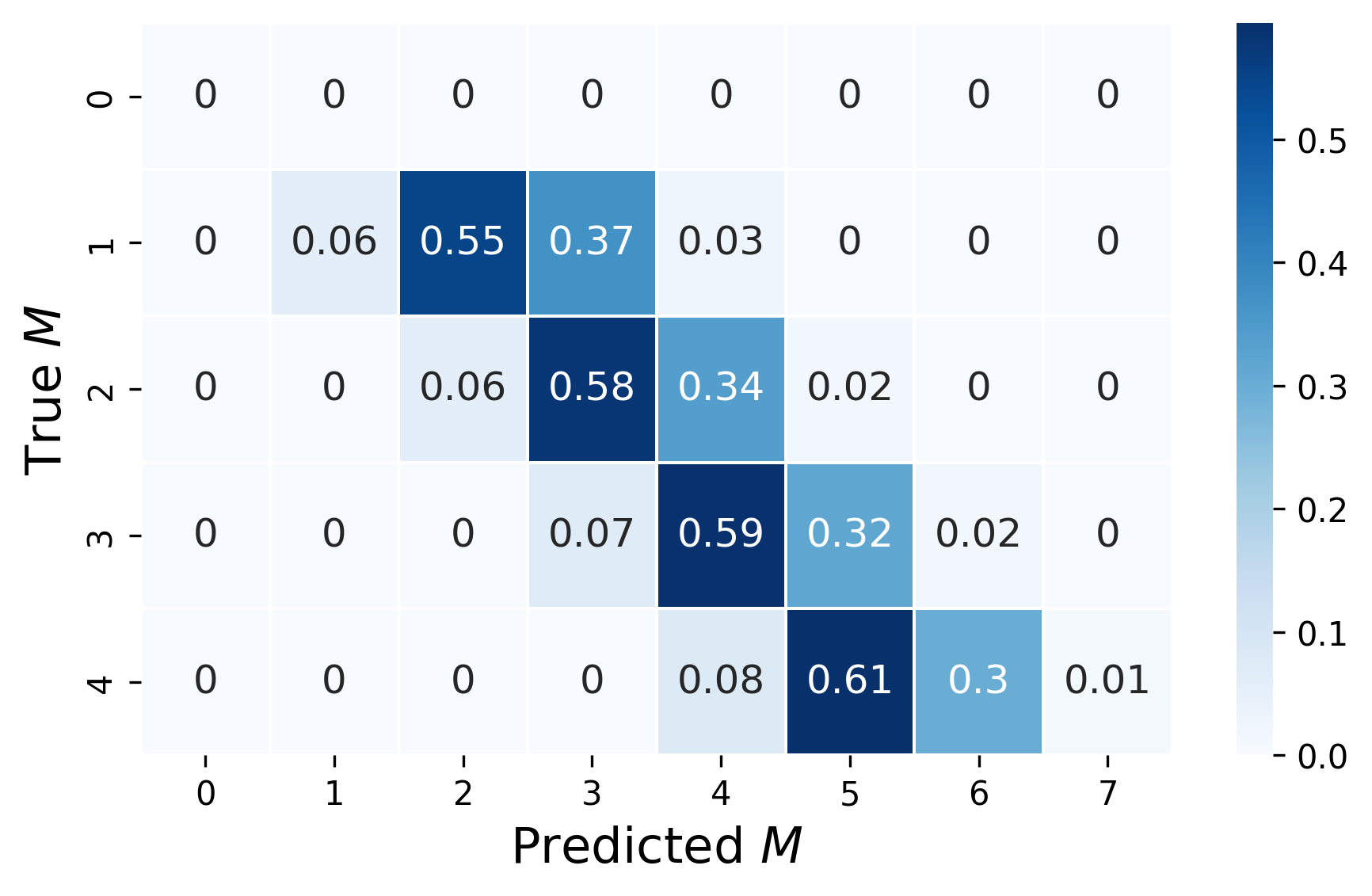}
    \label{subfig:source_enum_conf_mat_AIC}
  \end{subfigure}%
  \vspace*{\fill}   % maximize separation between the subfigures
  
  \begin{subfigure}{0.35\textwidth}
    \caption{DAFC}
    \includegraphics[width=\linewidth]{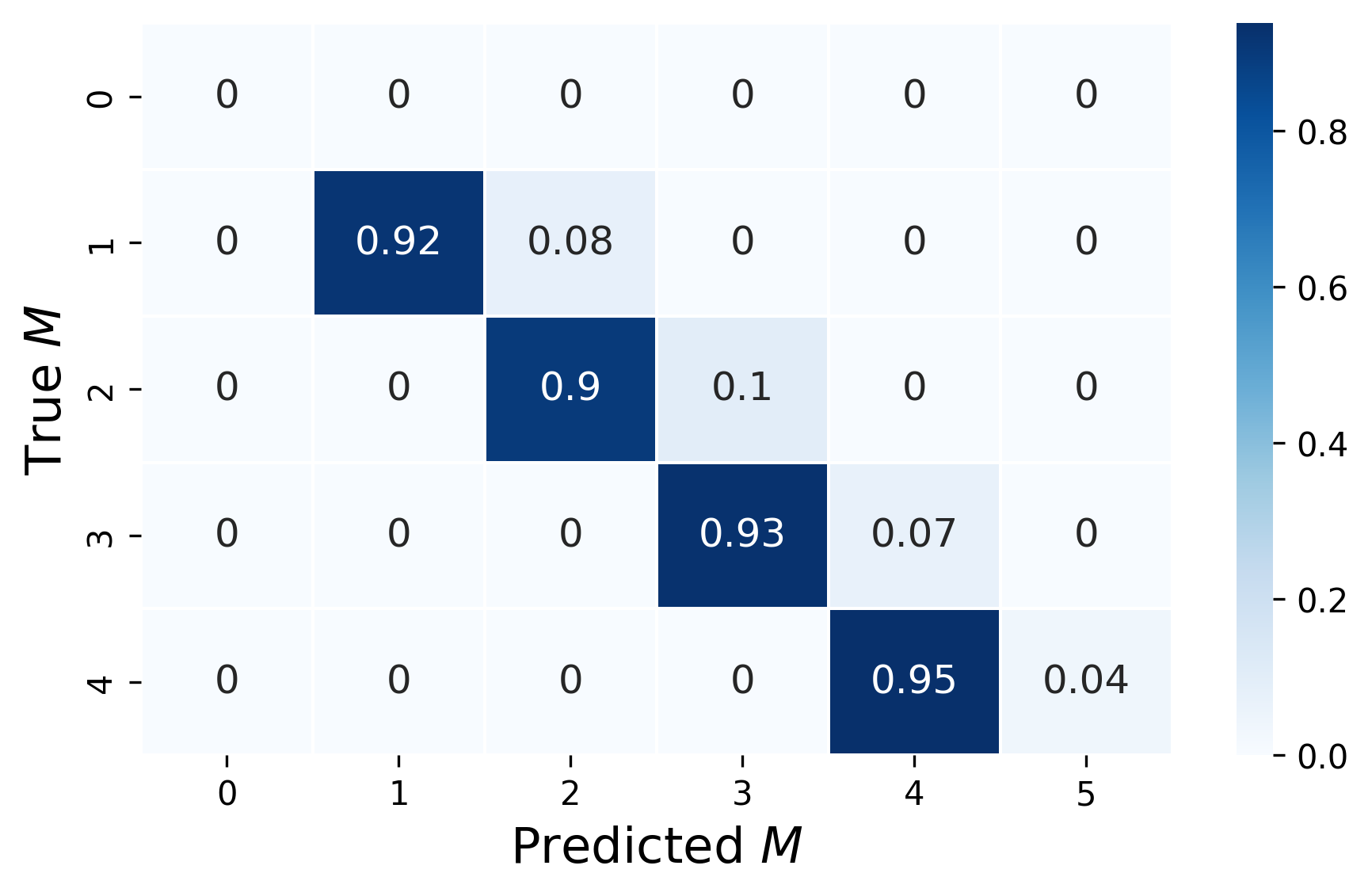}
    \label{subfig:source_enum_conf_mat_DAFC}
  \end{subfigure}%
    \caption{Confusion matrix for source enumeration, $\text{SIR}=0\;dB$, sources located at \textcolor{black}{$\{\theta_1,\theta_2,\theta_3,\theta_4\}=\theta_c + \{-2\Delta\theta, -\Delta\theta, \Delta\theta, 2\Delta\theta\}$, with $\theta_c\sim\text{U}\left(-15^\circ,15^\circ\right)$, and $\Delta\theta=15\degree$}. (a) MDL, (b) AIC, (c) proposed DAFC-based NN.} 
\label{fig:source_enum}
\end{figure}

\subsubsection{Multiple Source Enumeration}

The source enumeration performance is evaluated in this experiment.
The DOAs of the sources are selected from the set of following values: \textcolor{black}{$\{\theta_1,\theta_2,\theta_3,\theta_4\}=\theta_c + \{\Delta\theta, -\Delta\theta, -2\Delta\theta, 2\Delta\theta\}$} such that for $M$ sources, the DOAs are selected to be the first $M$ DOAs.
The interference is located at \textcolor{black}{$\theta_c\sim\text{U}\left(-15^\circ,15^\circ\right)$}.
The proposed DAFC-based NN approach is compared to the MDL and AIC~\cite{wax1985detection}.
Fig.~\ref{fig:source_enum} shows the source enumeration confusion matrices for the MDL, AIC, and the proposed DAFC-based NN with $\text{SIR}=0\;dB$. 

Figs. \ref{subfig:source_enum_conf_mat_MDL} and    \ref{subfig:source_enum_conf_mat_AIC} show that in both the MDL and the AIC, the predicted number of sources has a constant bias for each true $M$ due to the spatially-colored interference.
Fig.~\ref{subfig:source_enum_conf_mat_DAFC} shows the source enumeration performance of the proposed DAFC-based NN approach in the presence of spatially colored interference. 
The DAFC-based NN identifies the interference and does not count it as one of the sources by outputting a low probability for angular grid points near $\theta_c$, resulting in a better source enumeration performance.

\subsubsection{Loss Weights}

This experiment evaluates the effect of the loss weight update factors, $\{\beta^{(l)}\}_{l=1}^{N_w}$, introduced in \eqref{eq:loss_weight_decay}, on the confidence level in the spatial spectrum.
Let $\widetilde{\mathbf{B}}$ denote the set of $\{\beta^{(l)}\}_{l=1}^{N_w}$ values used in the proposed approach.
The loss weights, $\{w_i^{(t)}\}_{i=1}^d$, are defined by the factors $e_0^{(t)},e_1^{(t)}$ according to \eqref{eq:loss_weights_def}, and are introduced to provide a trade-off between the penalty obtained on source/interference and the penalty obtained for the rest of the output spatial spectrum.

For comparison, we set $\mathbf{B}_0=\{10^{-6}, 3.98\cdot10^{-6}, 1.58\cdot10^{-5}, 6.31\cdot10^{-5}, 2.51\cdot10^{-4}, 10^{-3}\}$, and $\mathbf{B}_1=\{10^{-3}, 3.98\cdot10{-3}, 0.0158, 0.063, 0.25, 0.1\}$ as two sets of loss weight update factors.
For $\mathbf{B}_0$, the loss weight update factors are closer to $0$. Hence the loss weights emphasize the source/interference, since $e_1^{(t)}\ll e_0^{(t)}$ which, according to to~\eqref{eq:loss_weights_def}, translates to larger $w_i^{(t)}$ for source/interference grid points.
For $\mathbf{B}_1$ the values are closer to $1$. Hence the loss weights are more equally distributed among grid points since $e_1^{(t)}\approx e_0^{(t)}$.
The experiment in \ref{subsubsec:eval_1_source} is repeated here for the DAFC-based NN approach with the two additional $\mathbf{B}_0,\mathbf{B}_1$ values mentioned above.

\begin{figure}[ht]
    \centering
    \includegraphics[width=0.49\textwidth]{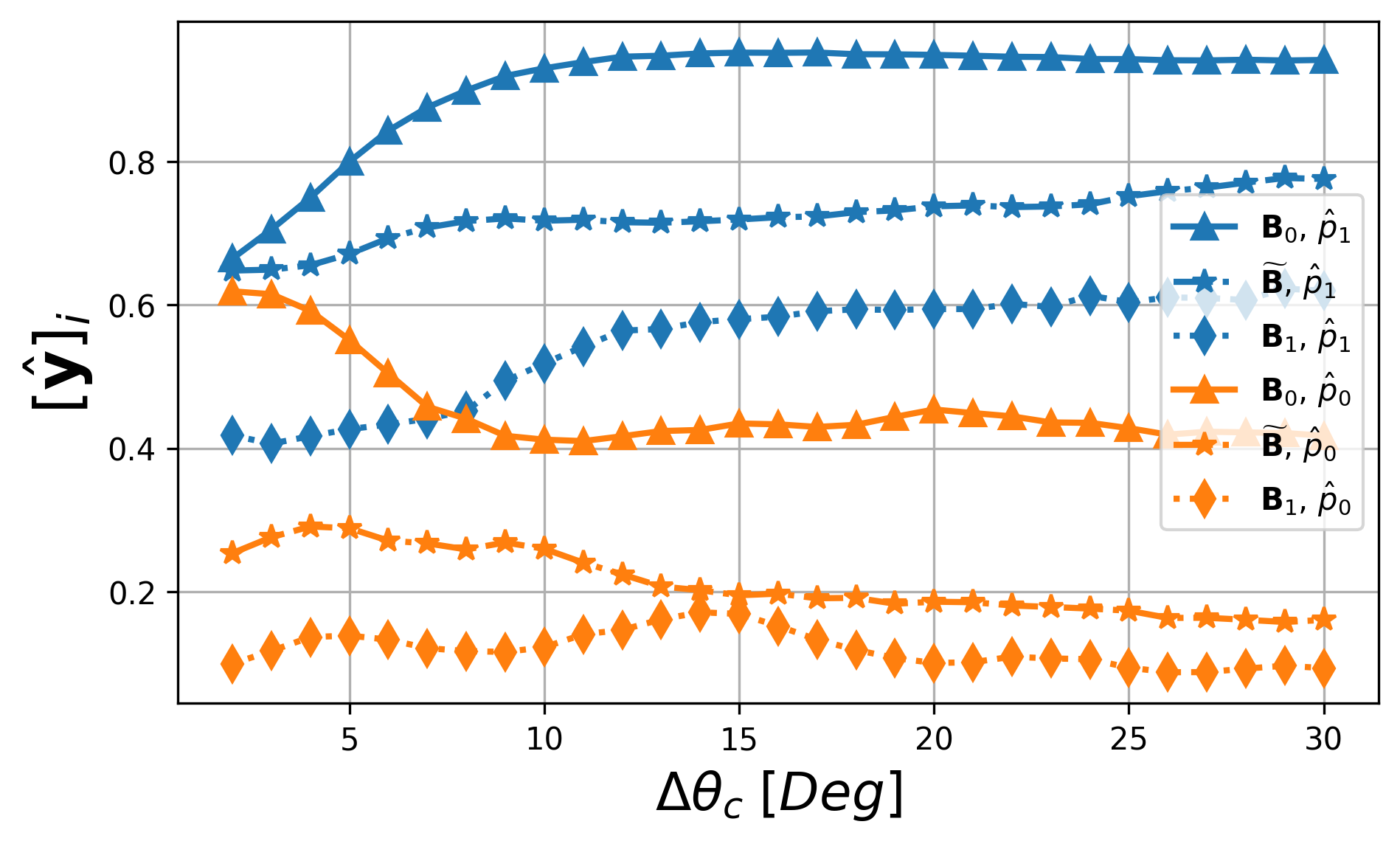}
    \caption{Loss weight update factor impact on probability levels obtained in the DAFC-based NN's spatial spectrum, single target at $\theta_1=0.55\degree$ with interference at $\theta_c=\theta_1+\Delta\theta_c$, $\text{SIR}=-5\;dB$. 
    $\hat{p}_1$ represents the probability obtained for source-containing grid points. $\hat{p}_0$ represents the probability obtained for non-source grid points.} 
    \label{fig:target_score}
\end{figure}

Let $\hat{p}_1$ represent the probability assigned for the source-containing grid point in the estimated label $\hat{\mathbf{y}}$.
Let $\hat{p}_0$ represent the maximum over probabilities assigned for non-source grid points in $\hat{\mathbf{y}}$, excluding a $5$-grid point guard interval around the source.
Fig.~\ref{fig:target_score} shows $\hat{p}_1$ and $\hat{p}_0$ for various angular separations between the source and interference for $\text{SIR}=-5\;dB$. 
For $\mathbf{B}_0$, the source's contribution to the loss value is substantially higher, which results in a higher probability for the source-containing grid point.
However, this results in a higher probability obtained for non-source grid points since their contribution to the loss value is negligible compared to the source-containing grid point, increasing ``false-alarm'' peaks in the spatial spectrum, subsequently increasing the estimation error.
Correspondingly, for $\mathbf{B}_1$ the source's contribution to the loss value is less significant, which results in a low probability assigned for the source-containing grid points, as well as a low probability for non-source grid points.

\section{Conclusion}\label{sec:conclusion}

This work addresses the problem of DOA estimation and source enumeration of an unknown number of sources within heavy-tailed, non-Gaussian, and spatially colored interference. A novel DAFC-based NN approach is proposed for this problem. 
The DAFC mechanism applies a structured transformation capable of exploiting the interference non-Gaussianity for its mitigation while retaining a low complexity of learnable parameters.
The proposed DAFC-based NN approach is optimized to provide an interference-mitigated spatial spectrum using a loss weight scheduling routine, performing DOA estimation and source enumeration using a unified NN.

The performance of the proposed approach is compared to MVDR, CNN-based, and FC-based approaches.
Simulations showed the superiority of the proposed DAFC-based NN approach in terms of probability of resolution and estimation accuracy, evaluated by RMSD, especially in weak signal power, a small number of snapshots, and near-interference scenarios.
The source enumeration performance of the proposed DAFC-based NN approach was compared to the  MDL and AIC. It was shown that in the considered scenarios, the proposed approach outperforms the MDL and the AIC in the source enumeration accuracy.

\bibliographystyle{IEEEtran}
\bibliography{main}

\begin{IEEEbiography}[{\includegraphics[width=1in,height=1.25in,clip,keepaspectratio]{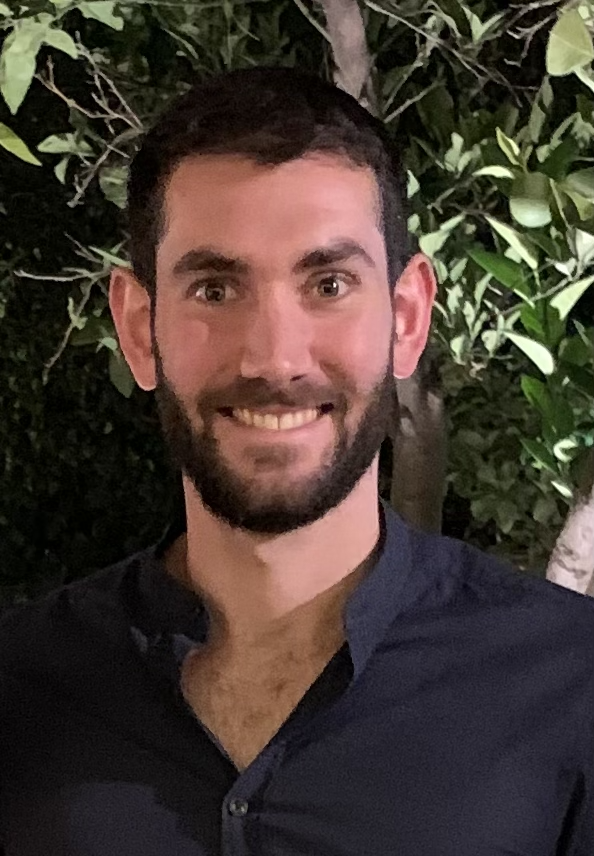}}]%
{Stefan Feintuch} received B.sc. and M.Sc. in electrical and computer engineering from the Ben-Gurion University of the Negev, Beer Sheva, Israel, in 2022 and 2023, respectively.
His research interests include machine learning, deep learning, radars, and signal processing.
\end{IEEEbiography}

\begin{IEEEbiography}[{\includegraphics[width=1in,height=1.25in,clip,keepaspectratio]{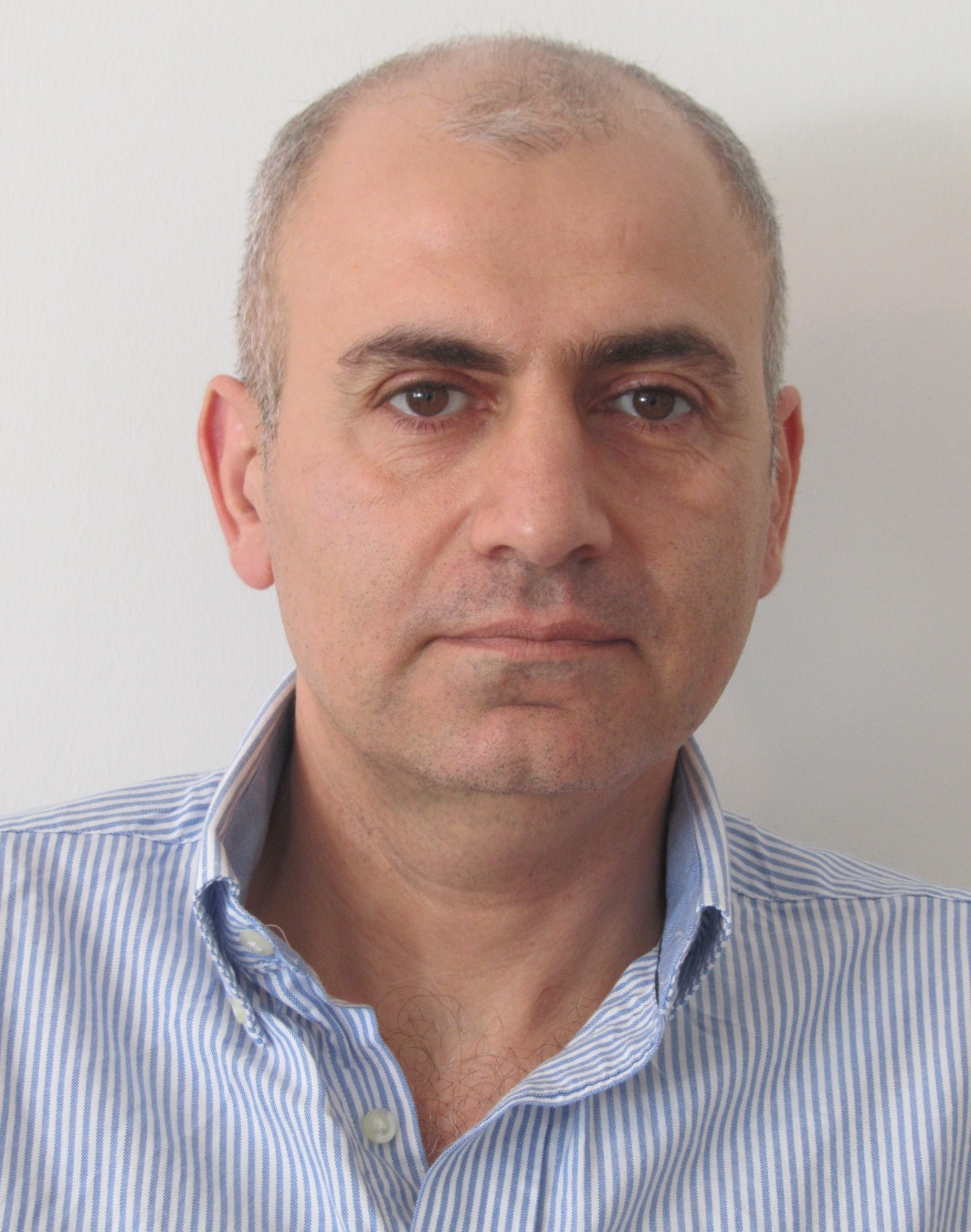}}]%
{Joseph Tabrikian} (Fellow, IEEE) received the B.Sc., M.Sc., and Ph.D. degrees in Electrical Engineering from the Tel-Aviv University, Tel-Aviv, Israel, in 1986, 1992, and 1997, respectively. During 1996–1998 he was with the Department of Electrical and Computer Engineering (ECE), Duke University, Durham, NC as an Assistant Research Professor. In 1998, he joined the Department of ECE, Ben-Gurion University of the Negev, Beer-Sheva, Israel, and served as the department head during 2017-2019. In May 2019 he established the school of ECE and served as its head till August 2021. He served as an Associate Editor (AE) for the IEEE Transactions on Signal Processing during 2001–2004 and 2011-2015, and is currently a Senior Area Editor (SAE) for these transactions. He served as AE and SAE of the IEEE Signal Processing Letters during 2012-2015 and 2015-2018, respectively. He was a member of the IEEE Sensor Array and Multichannel (SAM) technical committee during 2010-2015 and was the technical program co-chair of the IEEE SAM 2010 workshop. During 2015-2021 he served as a member of the Signal Processing for Multisensor Systems (SPMuS) of EURASIP and during 2017-2022 he was a member IEEE SPTM technical committee. He is co-author of 7 award-winning papers in IEEE conferences and workshops. His research interests include estimation and detection theory, learning algorithms, and radar signal processing. 
\end{IEEEbiography}

\begin{IEEEbiography}[{\includegraphics[width=1in,height=1.25in,clip,keepaspectratio]{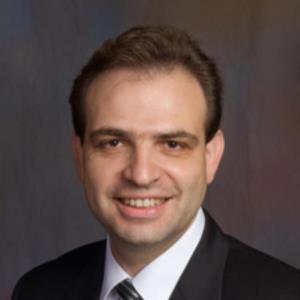}}]%
{Igal Bilik}(S'03-M'06-SM'21) received B.Sc., M.Sc., and Ph.D. degrees in electrical and computer engineering from the Ben-Gurion University of the Negev, Beer Sheva, Israel, in 1997, 2003, and 2006, respectively. During 2006–2008, he was a postdoctoral research associate in the Department of Electrical and Computer Engineering at Duke University, Durham, NC. During 2008-2011, he has been an Assistant Professor in the Department of Electrical and Computer Engineering at the University of Massachusetts, Dartmouth. During 2011-2019, he was a Staff Researcher at GM Advanced Technical Center, Israel, leading automotive radar technology development. Between 2019-2020 he was leading Smart Sensing and Vision Group at GM R\&D, where he was responsible on development state-of-art automotive radar, lidar and computer vision technologies. Since Oct. 2020, Dr. Bilik is an Assistant Professor in the School of Electrical and Computer Engineering at the Ben-Gurion University of the Negev. Since 2020, he is a member of IEEE AESS Radar Systems Panel and a vice-Chair of Civilian Radar Committee. Dr. Bilik is an Acting Officer of IEEE Vehicular Technology Chapter, Israel. Dr. Bilik  has more than 170 patent inventions, authored more than 60 peer-reviewed academic publications, received the Best Student Paper Awards at IEEE RADAR 2005 and IEEE RADAR 2006 Conferences, Student Paper Award in the 2006 IEEE 24th Convention of Electrical and Electronics Engineers in Israel, and the GM Product Excellence Recognition in 2017.
\end{IEEEbiography}

\begin{IEEEbiography}
[{\includegraphics[width=1in,height=1.25in,clip,keepaspectratio]{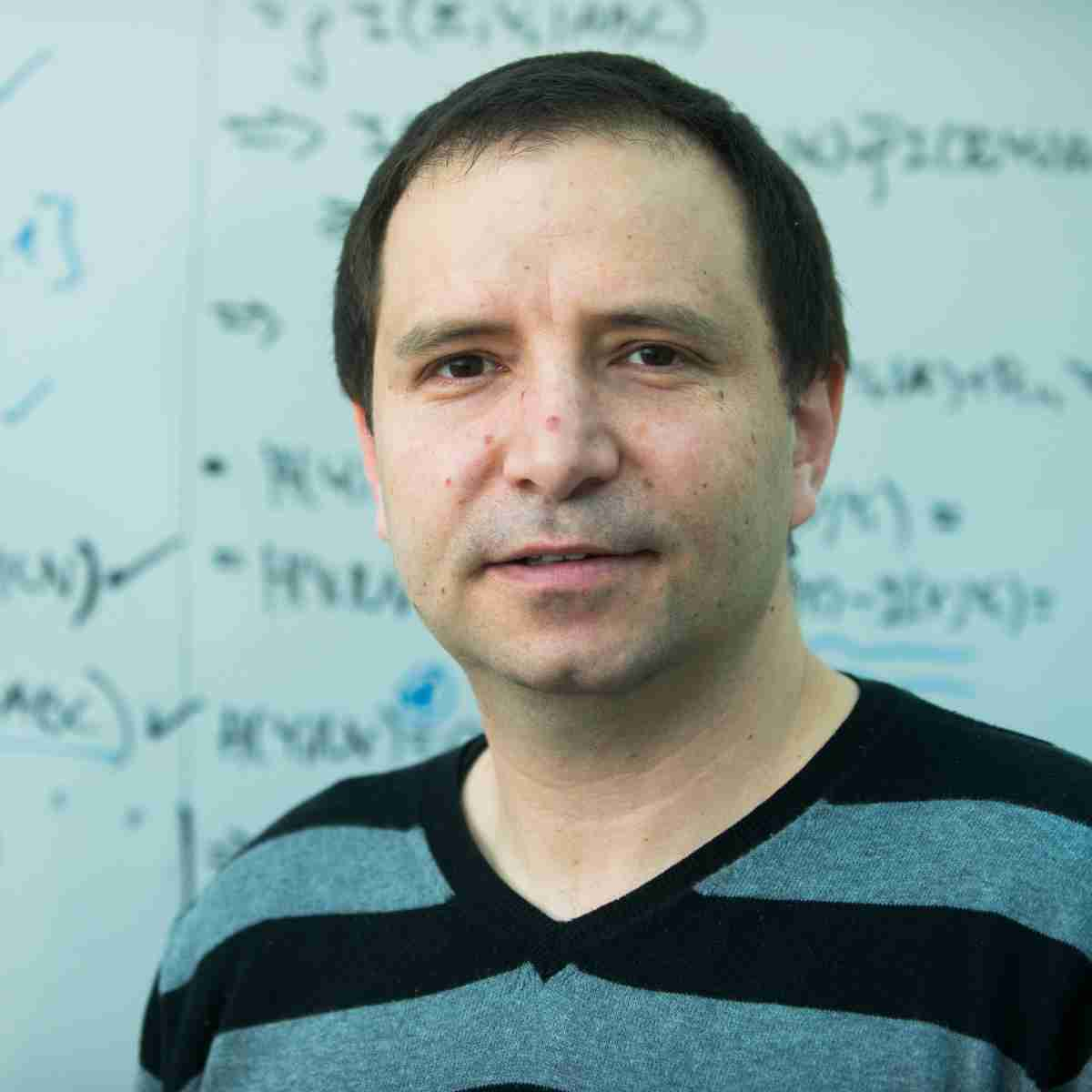}}]%
{Haim H. Permuter} (M'08-SM'13) received his B.Sc.\@ (summa cum laude) and M.Sc.\@(summa cum laude) degrees in Electrical and Computer Engineering from the Ben-Gurion University, Israel, in 1997 and 2003, respectively, and the Ph.D. degree in Electrical  Engineering from Stanford University, California in 2008. Between 1997 and 2004, he was an officer at a research and development unit of the Israeli Defense Forces. Since 2009 he is with the department of Electrical and Computer Engineering at Ben-Gurion University where he is currently  a professor, Luck-Hille Chair in Electrical Engineering. Haim also serves as head of the communication,cyber/ and information track in his department. Prof. Permuter is a recipient of several awards, among them the Fullbright Fellowship, the Stanford Graduate Fellowship (SGF), Allon Fellowship, and and the U.S.-Israel Binational Science Foundation Bergmann Memorial Award. Haim served on the editorial boards of the IEEE Transactions on Information Theory in 2013-2016 and has been reappointed again in 2023.
\end{IEEEbiography}

\end{document}